\documentclass[universe,article,accept,moreauthors,pdftex]{mdpi}

\firstpage{1} 
\pubvolume{1}
\issuenum{1}
\articlenumber{0}
\pubyear{2021}
\copyrightyear{2020}
\datereceived{25 November 2021} 
\dateaccepted{21 December 2021} 
\datepublished{} 
\hreflink{https://doi.org/} 
\pdfoutput=1

\usepackage{tabularx}
\usepackage{graphicx}
\usepackage{adjustbox}
\usepackage[flushleft]{threeparttable}

\Title{A Long-term Study of  Ultraluminous X-ray Sources in NGC 891}

\TitleCitation{A Long-term Study of  Ultraluminous X-ray Sources in NGC 891}

\Author{Nicholas M. Earley$^{1,\dagger,}$*, Vikram V. Dwarkadas\orcidA{}$^{1,\dagger,}$* and Victoria Cirillo$^{2}$}

\AuthorNames{Nicholas M. Earley, Vikram V.  Dwarkadas and Victoria Cirillo}

\AuthorCitation{Earley, N.; Dwarkadas, V.; Cirillo, V.}

\address{%
$^{1}$ \quad Department of Astronomy and Astrophysics, University of
Chicago, 5640 S. Ellis Ave, Chicago, IL 60637, USA\\
$^{2}$ \quad Department of Physics and Engineering Physics, Fordham University, 441 E. Fordham Rd., Bronx, NY 10458, USA}

\corres{Correspondence: earleyn138@uchicago.edu (N.E.),  vikram@astro.uchicago.edu (V.D.)}

\firstnote{These authors contributed equally to this work.}

\abstract{We perform empirical fits to the \emph{Chandra} and \emph{XMM-Newton} spectra of three ultraluminous X-ray sources (ULXs) in the edge-on spiral galaxy NGC 891, monitoring the region over a seventeen year time window.  One of these sources has been visible since the early 1990s with \emph{ROSAT} and has been observed multiple times with \emph{Chandra} and \emph{XMM-Newton}. Another has been visible since 2011. We build upon prior analyses of these sources by analyzing all available data at all epochs. Where possible \emph{Chandra} data is used, since its superior spatial resolution allows for more effective isolation of the emission from each individual source, thus providing a better determination of their  spectral properties. We also identify a new transient ULX, CXOU J022230.1+421937, which faded from view over the course of a two month period from Nov 2016 to Jan 2017. Modeling of each source at every epoch was conducted using six different models ranging from thermal bremsstrahlung to accretion disk models. Unfortunately, but as is common with many ULXs, no single model yielded a much better fit than the others. The two known sources had unabsorbed luminosities that remained fairly consistent over five or more years. Various possibilities for the new transient ULX are explored.}

\keyword{X-rays; galaxies: individual (NGC 891); accretion, accretion disks} 

\begin{document}
\section{Introduction}
Ultraluminous X-ray sources (ULXs) are point-like sources found far from the nucleus of external galaxies, with X-ray luminosities that exceed $10^{39}$ erg s$^{-1}$  \citep{Kaaret17}. ULXs were first discovered in the 1980s by the \emph{Einstein Observatory}, although the spectral resolution of \emph{ROSAT}  was needed to study their spectral properties \citep{1981ApJ...246L..61L, 1989ARA&A..27...87F}. Over the past twenty years, \emph{Chandra} and \emph{XMM-Newton} have provided even greater insight into the spectral properties and potential identity of these sources \cite{fs11}. ULXs were initially thought to be intermediate mass black holes (IMBHs) with masses between $10^2 - 10^6 \textrm{ M}_\odot$, under the assumption that their emission and accretion rates were isotropic and Eddington-limited \citep{Colbert1999, 2001MNRAS.321L..29K, 2003ApJ...596L.171G}. If emission is instead arising from an accretion disk surrounding the compact object, which is presumably in a binary system, then the luminosty can exceed the Eddington limit \citep{ss73, Poutanen2007, Fabrika2021}.  The latter model, where a supercritical accretion disk forms, and accretion occurs in  super-Eddington mode, seems to be more favored nowadays \citep{Fabrika2021}.

Theoretical models, and population synthesis studies, also seem to suggest that the majority of ULXs could be accreting highly magnetized NSs \citep{Mushtukov2015, Kuranov2020}. \citet{Mushtukov2021} showed that pulsating ULXs were not strongly beamed, and therefore their observed luminosity must be close to their actual luminosity.  \citet{King2017} have argued that while only a small number of systems have shown pulsations till now, that is so because magnetic neutron stars will only emit observable pulses for a short fraction of their lifetime. They conclude that a much larger fraction of ULXs harbor neutron star accretors than the number of pulsating ULXs. It is  increasingly likely that the population of ULXs as a whole comprises mainly of super-Eddington accreting compact objects, with several being confirmed as neutron stars (NSs) since 2014 \citep{Bachetti2014, Walton2014, Middleton2011, Mukherjee2015, Pinto2021}. Others could be super-Eddington accreting stellar-mass BHs, while the brightest could still potentially be IMBHs \citep{gladstone13, Fabrika2021}.

The X-ray spectra of ULXs show a curvature, with a sharp turnover above 5 keV and residuals below 1 keV \citep{Stobbart2006, Walton2014}. A variety of models have been used to fit the spectra and explore the physical nature of the systems \citep{gladstone11}. The spectral features appear to be consistent with accretion onto highly magnetized NSs. \citet{Mushtukov2017}, for example, found that when the accretion rates were high, the resulting multicolor blackbody spectrum originating from the magnetospheric surface was suggestive of observed ULX spectra.  Unfortunately it is quite common to find degeneracy between spectral models suggestive of different physical processes \citep{Koliopanos2017}. The latter suggest that much deeper broadband spectra, and time resolved spectroscopy are needed to resolve discrepancies between models. Accurate spectral modeling is thus a critical first step in assessing emission mechanisms of ULXs.

Many ULXs show persistent large luminosities over several years, or even decades as we show herein. Others exhibit short timescale variability and can disappear over the course of months \citep{2009ApJ...702.1679K, 2013MNRAS.433.1023G, Middleton2012, 2013ApJ...775...21B}. Strong variability in ULX luminosity could be attributable in some cases to NSs entering the ``propeller" regime of accretion, in which accretion is cut off at the pulsar's magnetospheric radius \citep{1975A&A....39..185I, 1986ApJ...308..669S, Tsygankov2016, Erkut2019}. Observation of the propeller effect could allow a determination of the source magnetic field. Alternatively, transient sources have been proposed to be X-ray binaries such as low mass X-ray binaries (LMXBs) seen during outbursts \citep{2018MNRAS.476.4272E, Middleton2012}.

Long-term monitoring of ULXs using a wide range of spectral models is required in order to  fully determine and understand the nature of these sources. In this paper, we model the spectra of three ULXs in the edge-on, barred spiral galaxy NGC 891 using archival \emph{XMM-Newton} and \emph{Chandra} observations dating since 2000. NGC 891 is a nearby analog to the Milky Way in terms of luminosity, color, and morphology \citep{1991rc3..book.....D, 1981A&A....95..116V, 1995A&A...299..657G}. This particular region of interest has been extensively observed because it hosts the supernova SN 1986J. Various ULXs  have appeared around the SN over time. ULX-1 (source coordinates 02:22:31.36 +42:20:24.4 \citep{2008ApJ...687..471B}; Figure 1) has been visible the longest, dating since the early 1990s when it was first observed with \emph{ROSAT} \citep{1998ApJ...493..431H}. ULX-2 (3XMM J022233.4+422026) was first discovered in a 2011 \emph{XMM-Newton} observation by \citet{hketal12} (hereafter HK12). The spectral modeling presented by HK12 was based solely on  \emph{XMM-Newton} observations; however, contamination of the ULX-2 \emph{XMM-Newton} spectrum from the  nearby ULX-1 (within about 25"), as well as SN 1986J, is possible. HK12 used 6 different spectral models, characteristic of ULXs, to fit the data; we adopt this same set of models in our comprehensive study of all datasets. In this work we also report the discovery of a new transient source detected in our 2016 \emph{Chandra} observation, CXOU J022230.1+421937, whose name follows the naming convention for a newly discovered \emph{Chandra} object. It is hereafter referred to as ULX-3. Using subsequent \emph{XMM-Newton} observations, we show that this bright source decreased considerably in luminosity over the course of two months.

Of the 3 ULXs investigated in this work, ULX-2 appears to have been relatively well studied in the past \cite{hketal12, hketal18, dageetal21}. However, spectral fits to the 2016 \emph{Chandra} data have not been previously reported. ULX-1, although having been visible for over 3 decades, does not seem to have been investigated as thoroughly. While flux measurements from the \emph{Chandra} 2000 and 2011 data have been reported \citep{bergheaetal08, swartzetal11, dageetal21}, spectral fits using other \emph{Chandra} and \emph{XMM} datasets are not found in the literature. Except HK12, other authors generally used only 1 or 2 spectral models to fit the data. ULX-3 is a new transient which, so far as we can tell, has not been previously reported. Altogether, about half the available data appears to not have been investigated earlier.

Irrespective of whether they have been studied earlier or not, in this work we analyze all the available data on all the ULXs using the same spectral models as HK12. We present a comprehensive and systematic analysis of all available \emph{XMM-Newton} and \emph{Chandra} data at all epochs on each of the 3 ULXs. The long term monitoring of ULXs presented here, spanning a time baseline of almost two decades, at this level of spectral analysis, is quite rare. It represents a repository of data, all fitted using the same spectral models, that could prove extremely useful for further investigations or theoretical modeling. The \emph{Chandra} data is essential in isolating the emission from each source, given the crowding of sources. With this large, rich, but difficult to analyze dataset, due to the crowding of the sources, our goal in this paper is to carry out the comprehensive spectral modeling of all three objects, at all available epochs, using both \emph{Chandra} and \emph{XMM-Newton} data, with spectral fitting using the 6 models as used by HK12. In a followup paper, we will present an in-depth temporal analysis of the light curve of each source at every epoch, exploring the time variability and searching for pulsations and quasi-periodic oscillations. We will also explore multiwavelength observations of the newly discovered ULX-3 in a bid to determine the nature of this transient.

In Section 2, we describe the \emph{XMM-Newton} and \emph{Chandra} observations of the sources; in Section 3, we perform X-ray spectral fits using thermal, power-law, and accretion disk models; and in Section 4, we discuss the results of the modeling and estimates of the fluxes which confirm these sources are ultraluminous. We also discuss the nature of the potential transient. Conclusions are given in Section 5. Throughout the paper, we use $z=0.001763$ for the redshift of the galaxy \citep{2009MNRAS.398..722T}, adopting the luminosity distance as $d \approx 9$ Mpc \citep{10.1111/j.1365-2966.2005.09336.x}, with $H_0 = 71 \text{ km } \text{s}^{-1} \text{ Mpc}^{-1}$, $\Omega_\Lambda = 0.73$, $\Omega_m = 0.27$ \citep{2007ApJS..170..377S}.
 
\section{Data Reduction and Analysis}
The region containing the three ULXs and nearby supernova SN 1986J has been frequently observed since the advent of \emph{Chandra} and \emph{XMM-Newton}. It was observed four times between 2000 and 2016 with the ACIS-S instrument on \emph{Chandra} and seven times between 2002 and 2017 with EPIC on \emph{XMM-Newton}. While the earliest \emph{Chandra} observation of the region was made in 2000, ULX-1 has been visible since at least 1991 when it was referred to as ``XNorth'' in \emph{ROSAT} observations \citep{1998ApJ...493..431H}. ULX-2 was visible in observations taken from 2011-2017. We identified a new source, CXOU J022230.1+421937, (ULX-3), in a \emph{Chandra} Nov 2016 observation.
\end{paracol}
\begin{specialtable}[H]	
\caption{2000-2017 archival observations of NGC 891. \label{tab:observations}}
\widefigure
\tablesize{\small} 
\begin{threeparttable}
\begin{tabular}{c | c | c | c | c | c | c}
\hline\hline
\textbf{Obs. ID} & \textbf{Mission}	& \textbf{Date} & \textbf{Exposure$^a$ (ks)} & \multicolumn{3}{c}{\textbf{Data count rate$^a$ ($10^{-2}$ counts s$^{-1}$)}} \\
& & & & ULX-1 & ULX-2 & ULX-3 \\
\hline
794 & \emph{Chandra} & 2000 Nov 1 & 50.9 & 3.9 &  &  \\
0112280101 & \emph{XMM-Newton} & 2002 Aug 22 & 7.8/12.9/12.6 & 6.8/2.7/2.7 &  &  \\
4613 & \emph{Chandra} & 2003 Dec 10 & 118.9 & 2.6 &  &  \\
14376 & \emph{Chandra} & 2011 Dec 20 & 1.8 & 0.7 & 8.4 &  \\
0670950101 & \emph{XMM-Newton} & 2011 Aug 25  & 94.8/112.6/113.8 & 7.9/2.7/2.8 & 39.9/13.2/11.9 &  \\
19297 & \emph{Chandra} & 2016 Nov 14 & 39.5 & 2.0 & 4.7 & 0.7 \\
0780760101 & \emph{XMM-Newton} & 2017 Jan 27 & 28.3/40.5/41.2 & 6.8/2.3/1.7 & 13.9/4.5/4.5 & 1.2/0.4/0.4 \\
0780760201 & \emph{XMM-Newton} & 2017 Jan 29 & 32.1/49.2/49.4 &  7.0/2.4/1.8 &  13.2/4.2/4.1 & 1.0/0.3/0.4 \\
0780760401 & \emph{XMM-Newton} & 2017 Feb 19 & 32.6/46.6/46.7 & 6.9/2.4/2.1 & 10.3/3.2/3.2 &  1.0/0.3/0.3 \\
0780760301 & \emph{XMM-Newton} & 2017 Feb 23 & 31.3/41.1/41.4 &  6.9/2.4/2.1 &  12.5/3.9/4.1 &  1.2/0.3/0.3 \\
0780760501 & \emph{XMM-Newton} & 2017 Feb 25 & 42.2/63.6/64.0 &  6.6/2.3/2.0 & 12.2/3.8/3.7 &  0.9/0.2/0.2 \\
\hline\hline
\end{tabular}
\begin{tablenotes}
    \footnotesize
    \item $^a$For \emph{XMM-Newton}, exposure times and count rates are given as those for EPIC pn/MOS1/MOS2, after filtering out background flaring. 
    \end{tablenotes}
\end{threeparttable}
\end{specialtable}  
\begin{paracol}{2}
\switchcolumn

\subsection{XMM-Newton observations}
The region of interest was observed by \emph{XMM-Newton} in 2002, 2011, and 2017. Details of the observations are shown in Table~\ref{tab:observations}. The data are reduced using version 18.0.0 of the XMM-SAS software following standard data reduction procedures for EPIC-pn, MOS1, and MOS2 spectra \citep{sas2017}. PN data are processed using \texttt{epproc} and MOS data using \texttt{emproc}. Light curves for each observation for each camera are filtered by count rate thresholds in order to eliminate instances of flaring. A circular 15" radius region is used to extract source spectra, while a circular 41.5" radius region, away from any source, is used for the background. Observations with the pn camera are shown for each epoch in Figure~\ref{fig:xmm_obs}, with only the first of five 2017 observations being presented. 

In the 2011 \emph{XMM-Newton} observation, ULX-2 lies close to the edge of the chip. Only a small portion of the source extraction region encompasses the edge of the chip. Due to the high central brightness of the source, the contribution to the count rate from this edge is found to be essentially negligible. 

The  spatial  resolution of \emph{XMM-Newton} is not high enough to uniquely isolate the emission from any of the sources. ULX-1 and ULX-2 are only 24" away from each other, and ULX-3 is 24" away from SN 1986J. A 25" region in \emph{XMM-Newton} observations on the other hand encircles 80\% of the radiated energy. In order to minimize contamination from nearby sources, the centers of the 15" radius apertures are offset from the source coordinates. However, contamination of the emission of any source from a nearby source (ULX or SN) may always be present. For this reason, we rely more on the spectral fits with \emph{Chandra}, with its 1" spatial resolution and a 90\% encircled energy radius of 4".

\begin{figure}[H]
\centering
\includegraphics[width=13 cm]{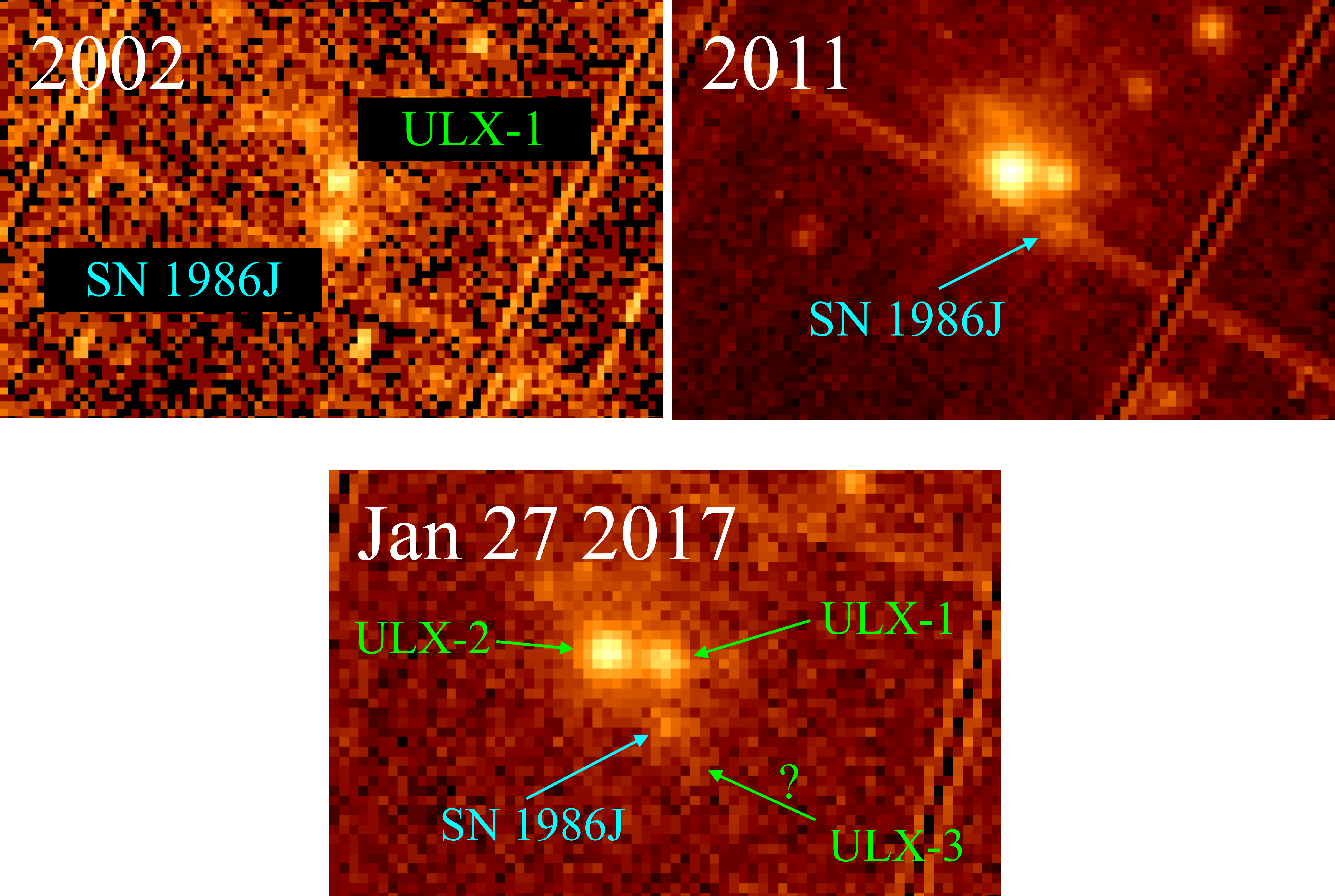}
\caption{2002, 2011, and Jan 27 2017 EPIC-pn observations in the 0.3-10.0 keV band. While only pn images are shown, pn and MOS spectra were simultaneously fit in the analysis. As can be seen, ULX-1 has been visible since 2000 and ULX-2 appeared in 2011. SN 1986J is in center of each image in between all of the sources. In the Jan 27 2017 EPIC-pn observation, the position denoting ULX-3 does not appear to harbor a source as bright or brighter than SN 1986J. \label{fig:xmm_obs}}
\end{figure}   

\subsection{Chandra observations}
The same region was observed with \emph{Chandra} ACIS-S in 2000, 2003, 2011, and 2016. The data are reduced and analyzed using the standard analysis pipeline in the CIAO software version 4.12 and CALDB 4.9.2.1 \citep{2006SPIE.6270E..1VF}. As seen in Figure~\ref{fig:chandra_obs}, emission from the various sources is mostly distinct and non-overlapping in contrast to the \emph{XMM-Newton} observations. However, the \emph{Chandra} observations also reveal that the source termed ULX-1 actually appears to be two sources in proximity to each other. A 1.5" radius aperture is used to isolate the ultraluminous source at each epoch, which is highlighted in the 2000 observation in Figure~\ref{fig:chandra_obs}. We note that it is not possible to isolate the ULX in the \emph{XMM-Newton} observations in this manner. For the other ULXs, a 4" radius aperture centered on the source is used as the source region, with the background region having a 30" radius offset from each individual source. Spectra are extracted using the \texttt{specextract} command. While the \emph{Chandra} observations in 2000, 2003, and 2016 had a large number of counts to provide high spectral resolution for all of the sources, the 2011 epoch had $<$ 2 ks of data, not enough to provide a sufficient number of counts for all sources. For ULX-1, with  three other epochs of high spectral resolution data available, we have chosen not to use the 2011 data since they are the least accurate. For ULX-2, however, with only one other epoch of \emph{Chandra} data available, results from 2011 are also presented. Spectra from ULX-1 are obtained in 2000, 2003, and 2016; from ULX-2 in 2011 and 2016; and ULX-3 in 2016 only. The observational information is available in Table~\ref{tab:observations}. 

\begin{figure}[H]
\centering
\includegraphics[width=13 cm]{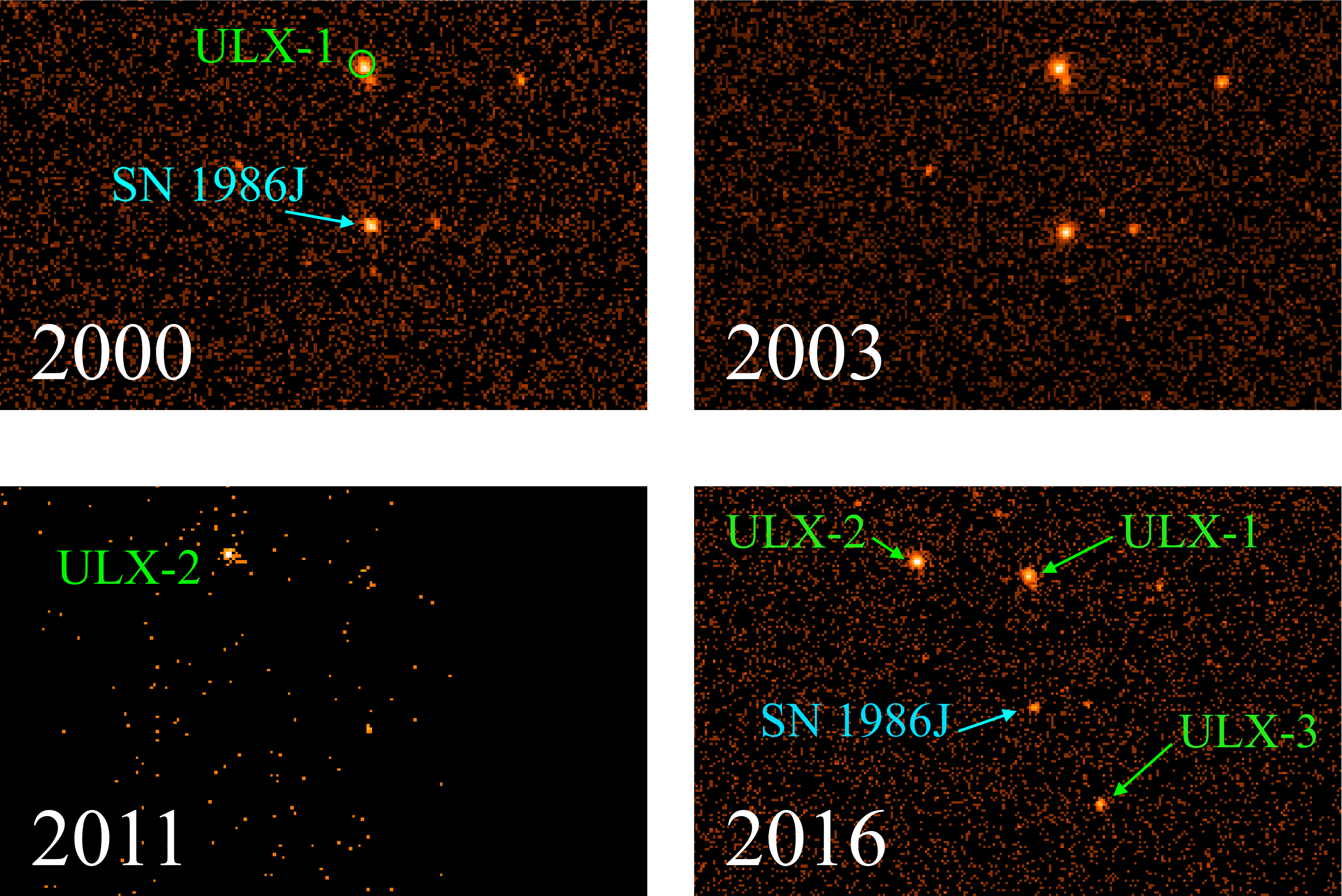}
\caption{2000, 2003, 2011, and Nov 2016 \emph{Chandra} observations in the 0.3-10.0 keV band. A 1.5" radius aperture denotes ULX-1 in the 2000 observation. The short exposure time in 2011 obscures the presence of SN 1986J and ULX-1. \label{fig:chandra_obs}}
\end{figure}



\section{Results}
The \emph{Chandra} and \emph{XMM-Newton} spectra are fit using Sherpa in CIAO version 4.12 \citep{2006SPIE.6270E..1VF}. Due to the high count numbers in nearly all epochs, spectra are background-subtracted and binned, and the \emph{chi2gehrels} statistic is employed in each fit. The flux and errors for each model are calculated using the Sherpa command \texttt{sample\_flux} with 1000 iterations. 1-sigma errors on best-estimate model parameter values are calculated when possible via the \texttt{covar} command, and with the \texttt{conf} command when \texttt{covar} fails. For \emph{XMM-Newton}, to ensure the tightest constraints on the model parameters, we simultaneously fit the EPIC-pn, MOS1, and MOS2 spectra. Given the stability of the spectra of ULX-1 and ULX-2 over time, the pn, MOS1, and MOS2 spectra taken over five observations in 2017 are merged for each camera, using the SAS task \texttt{epicspeccombine}. The resulting merged pn, MOS1, and MOS2 spectra are then fit simultaneously as for other epochs. 

ULX-2 was initially analyzed by HK12. Their empirical fits employed six different models appropriate to ULXs. In this paper, we fit the spectra of all the ULXs with the same models that they used in order to allow for direct comparison. HK12 solely considered the ULX-2 \emph{XMM-Newton} data from 2011, which as noted could suffer from contamination from nearby sources. \citet{hketal18} also fit the 2017 \emph{XMM-Newton} spectra. This paper expands upon their work by investigating all ULXs in this region over all available epochs, including \emph{Chandra} data which has the spatial resolution to isolate the emission from each source and reduce source contamination. All source models include a \texttt{tbabs} absorption component with abundances set to  \citet{2000ApJ...542..914W} prior to fitting. Thermal bremsstrahlung (\texttt{bremss}) and broken power-law (\texttt{bknpower}) models are some of the simpler models used, describing thermal continuum emission or non-thermal power-law emission that could denote synchrotron or inverse Compton processes for example. Given that the nature of ULXs is unknown, but that they may be compact objects and their emission may arise from a disk around a stellar mass black hole or IMBH, the following accretion disk models are also used: 

\begin{enumerate}
    \item The \textbf{\texttt{diskpbb}} model. This is referred to as a ``p-free'' multi-color disk blackbody (MCD) model in which the accretion disk temperature $T(r) \propto r^{-p}$. In the standard MCD model, each point of the disk is assumed to radiate like a blackbody with a temperature that scales as $T(r) \propto r^{-0.75}$ \citep{1984PASJ...36..741M}. In this model $p$ is allowed to vary, allowing for the possibility of advection-dominated disks ($p \sim 0.5$) in which photons cannot radiate away from the disk but are swept along with the accreting material \citep{Abramowicz1988}. In the case of super-Eddington accretion, this disk may become so optically thick that this advective process dominates.
    
    \item The two-component \textbf{\texttt{diskpbb}+\texttt{comptt}} model. This is an MCD disk plus Comptonization model in which the seed photons originating from the inner disk with a Wien spectrum are Comptonized in a surrounding corona \citep{2009MNRAS.397.1836G, 1994ApJ...434..570T}. 
    
    \item The \textbf{
    \texttt{diskir}} model. This is a model in which the inner regions of the disk are irradiated by Comptonized photons originating from the disk itself. Part of the resulting spectrum is due to the reflection of the Comptonized photons and the rest from the absorption and subsequent re-emission of such photons. This re-emission ultimately modifies the original disk emission had there not been any irradiation \citep{2009MNRAS.392.1106G}. 
    
    
    \item The \textbf{\texttt{kdblur2*reflionX}} model. This model describes emission arising from the reflection of radiated photons by an optically thick, constant density atmosphere \citep{1999MNRAS.306..461R, 2005MNRAS.358..211R}. The reflected spectrum is smoothed by the relativistic effects from the accretion disk surrounding a rotating black hole. The blurred reflection model is a table model that can be read into Sherpa using user-contributed scripts. 
\end{enumerate}

The results of the spectral fitting to the \emph{Chandra} and \emph{XMM-Newton} data along with the unabsorbed model fluxes in the 0.3-10.0 keV band are reported at each epoch in Tables~\ref{tab:chandra_spectra} and \ref{tab:xmm_spectra}, respectively. The 2017 \emph{XMM-Newton} spectral models for ULX-3 are included in Table~\ref{tab:ulx3_spectra}.
\newpage

\end{paracol}
\begin{specialtable}[H]	
\caption{\emph{Chandra} ULX spectral models. \label{tab:chandra_spectra}}
\widefigure
\begin{center}
\scalebox{0.73}{
\begin{threeparttable}
\begin{tabular}{c c c | c c c | c c | c}
\hline\hline
\textbf{Component} & \textbf{Parameter} & \textbf{Units} & \multicolumn{3}{c|}{\textbf{ULX-1}} & \multicolumn{2}{c|}{\textbf{ULX-2}} &
\textbf{ULX-3} \\
& & & 2000 & 2003 & 2016 & 2011$^*$ & 2016 & 2016 \\
\hline\hline
$\text{TB}_{\text{ABS}} \cdot \text{BREMSS}$ \\
\hline\hline
$\text{TB}_{\text{ABS}}$ & $N_{H}$ & $10^{22}$ cm$^{-2}$ & 
$0.87 \pm 0.06$ & 
$0.84 \pm 0.05$ & 
$0.87 \pm 0.16$ &
$0.17^{+0.15}_{-0.11}$ &
$0.16 \pm 0.04$ &
$2.20 \pm 0.16$ \\
$\text{BREMSS}$  & $kT$ & keV & 
$6.62 \pm 1.21$ & 
$6.31 \pm 0.88$ & 
$10.06 \pm 3.86$ & 
$10.12 \pm 5.66$ & 
$4.56 \pm 0.65$ & 
$24.12^{+a}_{-16.48}$ \\
$\chi^2$ &&& 
0.57 & 
0.71 & 
0.66 & 
0.75 &
0.80 &
0.67 \\
Absorbed ${F}_{X_{0.3-10.0\text{ keV}}}$ && $10^{-13}$ erg s$^{-1}$ cm$^{-2}$ &
$4.6 \pm 0.5$ &  
$2.8^{+0.2}_{-0.3}$ & 
$3.5^{+0.5}_{-0.6}$ &
$7.4^{+2.6}_{-2.9}$  &
$5.0^{+0.5}_{-0.6}$ &  
$1.4 \pm 0.1$\\
Unabsorbed ${F}_{X_{0.3-10.0\text{ keV}}}$ && $10^{-13}$ erg s$^{-1}$ cm$^{-2}$ &
$6.8^{+0.5}_{-0.6}$ &  
$4.1 \pm 0.3$ & 
$4.9^{+0.5}_{-0.6}$ &
$8.5^{+3.0}_{-3.3}$  &
$6.0^{+0.6}_{-0.5}$ &  
$2.1 \pm 0.1$\\
\hline\hline
$\text{TB}_{\text{ABS}} \cdot \text{BKNPOWER}$ \\
\hline\hline
$\text{TB}_{\text{ABS}}$ & $N_{H}$ & $10^{22}$ cm$^{-2}$ & 
$0.90 \pm 0.15$ &
$1.02 \pm 0.02$ &
$0.90 \pm 0.07$ &
$0.24^{+0.20}_{-0.15}$  &
$0.29 \pm 0.08$ & 
$1.65^{+0.92}_{-0.69}$ \\
BKNPOWER & $\Gamma_1$ && 
$1.50 \pm 0.37$ & 
$1.89 \pm 0.03$ &
$1.54 \pm 0.03$  &
$1.62^{+0.39}_{-0.33}$ &
$1.98 \pm 0.16$ &
$0.96^{+0.50}_{-0.46}$ \\
& $\Gamma_2$ && 
$1.88 \pm 0.12$ &
$1.77 \pm 0.41$ &
$9.0^{+a}_{-7.5}$ &
$6.45^a$ &
$2.14 \pm 0.46$ &
$4.70^{+a}_{-3.16}$\\
& BreakE & keV & 
$2.04 \pm 0.36$ &
$4.68^a$ &
$6.45 \pm 0.40$  &
$12.86^{+a}_{-12.83}$  &
$3.82 \pm 1.51$  &
$5.21^{+0.88}_{-2.08}$\\
$\chi^2 $ &&& 
0.55 & 
0.70 & 
0.69 & 
0.90 &
0.78 &
0.65 \\
Absorbed ${F}_{X_{0.3-10.0\text{ keV}}}$ && $10^{-13}$ erg s$^{-1}$ cm$^{-2}$ &
$4.8^{+2.1}_{-1.6}$ &  
$3.1^{+0.3}_{-0.2}$ & 
$3.2^{+0.3}_{-0.2}$ &
$8.1^{+5.1}_{-3.2}$  &
$5.2^{+1.5}_{-1.2}$ &  
$1.3^{+3.5}_{-0.9}$\\
Unabsorbed ${F}_{X_{0.3-10.0\text{ keV}}}$ && $10^{-13}$ erg s$^{-1}$ cm$^{-2}$ &
$7.3^{+2.3}_{-2.1}$ & 
$5.2^{+0.3}_{-0.2}$ &
$4.7^{+0.3}_{-0.2}$  &
$10.1^{+4.8}_{-3.5}$ &
$7.6^{+1.6}_{-1.3}$ &
$1.8^{+3.7}_{-1.1}$\\
\hline\hline
$\text{TB}_{\text{ABS}} \cdot \text{DISKPBB}$ \\
\hline\hline
$\text{TB}_{\text{ABS}}$ & $N_{H}$ & $10^{22}$ cm$^{-2}$ & 
$0.98 \pm 0.06$ & 
$0.97 \pm 0.06$ &
$0.60^{+0.32}_{-0.28}$ &
$0.20 \pm 0.18$ &
$0.28 \pm 0.05$ &
$1.45 \pm 0.81$\\
DISKPBB & $T_{\text{in}}$ & keV &
$3.99^{+a}_{-1.29}$ &
$4.32^{+a}_{-1.26}$ &
$1.94^{+1.11}_{-0.44}$ &
$3.16 \pm 1.63$ &
$3.87^{+a}_{-1.22}$ &
$2.04 \pm 0.84$\\
& $p$ &&
$0.533 \pm 0.008$ & 
$0.526 \pm 0.007$ & 
$0.73^{+0.55}_{-0.14}$ & 
$0.58 \pm 0.08$ &
$0.506 \pm 0.008$  & 
$1.0 \pm 0.8$\\
$\chi^2 $ &&& 
0.57 & 
0.69 & 
0.63 & 
0.82 &
0.76 &  
0.65 \\
Absorbed ${F}_{X_{0.3-10.0\text{ keV}}}$ && $10^{-13}$ erg s$^{-1}$ cm$^{-2}$ &
$5.8^{+4.4}_{-3.5}$ &  
$3.6^{+2.6}_{-2.3}$ & 
$3.2^{+8.4}_{-2.5}$ &
$5.8^{+31.3}_{-5.2}$ &
$5.4^{+4.8}_{-3.4}$ &  
$2.0^{+8.8}_{-1.8}$\\
Unabsorbed ${F}_{X_{0.3-10.0\text{ keV}}}$ && $10^{-13}$ erg s$^{-1}$ cm$^{-2}$ &
$9.4^{+7.1}_{-5.8}$ & 
$6.0^{+4.4}_{-3.8}$ & 
$4.0^{+10.9}_{-3.1}$ &
$6.7^{+37.1}_{-6.0}$ &
$7.7^{+7.2}_{-4.9}$ & 
$2.7^{+11.4}_{-2.5}$\\
\hline\hline
$\text{TB}_{\text{ABS}} \cdot (\text{DISKPBB}+\text{COMPTT})$ \\
\hline\hline
$\text{TB}_{\text{ABS}}$ & $N_{H}$ & $10^{22}$ cm$^{-2}$ & 
$0.98 \pm 0.06$ & 
$1.45^{+0.11}_{-0.64}$ &
$1.10^{+0.17}_{-0.33}$ &
$0.51^{+0.36}_{-0.38}$ &
$0.27^{+0.12}_{-0.02}$ &
$1.82^{+1.56}_{-1.79} $\\
DISKPBB & $T_{\text{in}}$ & keV & 
$1.31 \pm 0.53$ & 
$6.6 \pm 0.2$ &
$1.30^a$ &
$3.37 \pm 2.20$ & 
$9.84^{+a}_{-9.82}$ &
$0.47 \pm 9.4\mathrm{e}{-5}$ \\
& $p$ && 
$0.53$ (f) & 
$0.53$ (f) & 
$0.53$ (f) & 
$0.58$ (f) &
$0.51$ (f)  & 
$1$ (f) \\
COMPTT & $T0$ & keV &
$0.12 \pm 0.07$ &
$0.10 \pm 0.01$ &
$0.11^{+0.10}_{-a}$ &
$0.10^a$ &
$0.33^a$ &
$0.35^{+4.86}_{-a}$\\
& $kT_{e}$ & keV &
$15.8^a$ &
$47.8^a$ &
$41.4^{+625}_{-a}$ &
$33.9^{+47.4}_{-a}$ &
$50.4^a$ &
$50.7^a$ \\
& $\tau_p$ &&
$8.1^{+60.2}_{-a}$ &
$0.01^{+2.02}_{-a}$ &
$1.1^{+a}_{-0.7}$ &
$0.01^a$ &
$1.25^a$ &
$2.57 \pm 2.26$\\
$\chi^2 $ &&& 
0.57 & 
0.74 & 
0.89 & 
1.09 &
0.82 & 
0.96 \\
Absorbed ${F}_{X_{0.3-10.0\text{ keV}}}$ && $10^{-13}$ erg s$^{-1}$ cm$^{-2}$ &
$5.2^b$ &  
$3.0^b$ & 
$3.8^b$ &
$7.7^b$  &
$5.9^{+4.6}_{-3.6}$ &  
$0.6^{+1.6}_{-0.5}$\\
Unabsorbed ${F}_{X_{0.3-10.0\text{ keV}}}$ && $10^{-13}$ erg s$^{-1}$ cm$^{-2}$ &
$5.2^b$ &
$3.0^b$ &
$3.8^b$ &
$7.7^b$ &
$8.3^{+6.4}_{-5.0}$ &
$0.7^{+1.8}_{-0.5}$\\
\hline\hline
$\text{TB}_{\text{ABS}} \cdot \text{DISKIR}$ \\
\hline\hline
$\text{TB}_{\text{ABS}}$ & $N_{H}$ & $10^{22}$ cm$^{-2}$ &
$0.82 \pm 0.06$ &
$0.96 \pm 0.02$ &
$0.79 \pm 0.02$ &
$0.12^{+0.32}_{-0.10}$ &
$0.22 \pm 0.02$ &
$2.82 \pm 0.22$ \\
DISKIR & $kT_{\text{disk}}$ & keV & 
$0.82^{+0.46}_{-0.31}$ &
$0.55 \pm 0.03$ &
$1.04 \pm 0.23$ &
$1.13^{+a}_{-0.19}$ &
$0.93 \pm 0.22$ &
$0.17^{+0.12}_{-0.10}$\\
& $\Gamma$ &&
$1.73 \pm 0.61$ &
$1.96 \pm 0.07$ &
$2.0$ (f) &
$1.01^a$ &
$2.02 \pm 0.72$ &
$1.53 \pm 0.20$\\
& $L_c/L_d$ && 
$1.54 \pm 0.59$ &
$3.1^{+a}_{-2.1}$ &
$1.42 \pm 0.62$ &
$4.81^{+a}_{-3.89}$ &
$8.48^{+a}_{-7.35}$ &
$10.0^a$\\
& $kT_{e}$ & keV &
$5.04 \pm 1.17$ &
$82.8 \pm 4.1$ &
$5.17 \pm 0.09$ &
$5.01^a$ & 
$17.85 \pm 0.55$ &
$142.2 \pm 33.9$ \\
& $f_{\text{in}}$ &&
$0.1$ (f) & 
$0.1$ (f) & 
$0.1$ (f) & 
$0.1$ (f) & 
$0.1$ (f) &
$0.15$ (f) \\
& $f_{\text{out}}$ && 
$2.0\mathrm{e}{-5}^{+0.115}_{-a}$ &
$0.02 \pm 0.01$ &
$3.0\mathrm{e}{-5}^{+0.04}_{-a}$ & 
$0.002^a$ & 
$0.03 \pm 0.01$ &
$4.2\mathrm{e}{-6}^a$\\
& $r_{\text{irr}}$ & $r_{\text{in}}$ &
$1.01^{+1.00}_{-0.01}$ & 
$1.03^{+0.60}_{-a}$ & 
$1.01$ (f) & 
$1.0001$ (f) & 
$1.0 \pm 1.1\mathrm{e}{-5}$ &
$3.21^a$\\
& $\log r_{\text{out}}$ & $\log r_{\text{in}}$ & 
$3.84 \pm 2.35$ & 
$3.31 \pm 0.95$ & 
$3.5$ (f) & 
$5$ (f) & 
$5$ (f) &
$5$ (f)\\
$\chi^2 $ &&& 
0.60 & 
0.75 & 
0.77 & 
1.18 &
0.83 & 
0.995 \\
Absorbed ${F}_{X_{0.3-10.0\text{ keV}}}$ && $10^{-13}$ erg s$^{-1}$ cm$^{-2}$ &
$5.0^b$ &  
$3.2^{+1.3}_{-0.8}$ & 
$4.1^{+6.6}_{-3.2}$ &
$8.9^b$  &
$5.3^{+10.0}_{-3.8}$ &  
$1.4^b$\\
Unabsorbed ${F}_{X_{0.3-10.0\text{ keV}}}$ && $10^{-13}$ erg s$^{-1}$ cm$^{-2}$ &
$5.0^b$ &
$5.1^{+2.0}_{-1.3}$ &
$5.6^{+8.3}_{-4.4}$ &
$8.9^b$ &
$6.6^{+13.3}_{-4.7}$ &
$1.4^b$ \\
\hline\hline
$\text{TB}_{\text{ABS}} \cdot \text{KDBLUR2} \cdot \text{REFLIONX}$ \\
\hline\hline
$\text{TB}_{\text{ABS}}$ & $N_{H}$ & $10^{22}$ cm$^{-2}$ &
$1.30^{+0.17}_{-0.12}$ &
$1.35 \pm 0.10$ &
$1.18^{+1.31}_{-0.28}$ &
$1.35 \pm 0.32$ &
$0.63 \pm 0.11$ &
$2.48^{+1.92}_{-0.47}$ \\
KDBLUR2 & $R_{\text{in}}$ & $R_G$ &
$9.5^{+5.1}_{-a}$ &
$8.2 \pm 0.2$ &
$1.98^{+5.13}_{-a}$ &
$2.37 \pm 0.47$ &
$1.24^{+6.93}_{-a}$ &
$1.24^{+8.17}_{-a}$ \\
& $R_{\text{out}}$ & $R_G$ &
$400$ (f) &
$400$ (f) &
$400$ (f) &
$400$ (f) &
$400$ (f) &
$400$ (f) \\
& $i$ & deg & 
$85.3^{+a}_{-41.2}$ &
$55.8 \pm 8.6$ &
$84.5^{+a}_{-9.2}$ &
$31.9 \pm 15.8$ &
$52.9 \pm 7.0$ &
$90.0^{+a}_{-38.7} $\\
& $q_{\text{in}}$ &&
$3.70^{+a}_{-4.46}$ &
$6.31^{+a}_{-5.87}$ &
$5.21^{+a}_{-3.39}$ &
$8.94^{+a}_{-6.52}$ &
$5.32 \pm 0.59$ &
$9.46^{+a}_{-16.3}$\\
& $q_{\text{out}}$ && 
$3.0$ (f) & 
$3.0$ (f) & 
$3.0$ (f) &
$3.0$ (f) &
$3.0$ (f) &
$3.0$  (f) \\
& $R_{\text{break}}$ & $R_G$ &
$20.0$ (f) & 
$20.0$ (f) &
$20.0$ (f) &
$20.0$ (f) &
$20.0$ (f) &
$20.0$ (f) \\
REFLIONX & $\Gamma$ &&
$1.70^{+0.54}_{-a}$ & 
$1.85 \pm 0.15$ &
$1.80^{+0.48}_{-a}$ &
$1.70^{+0.81}_{-a}$ &
$1.74^{+0.34}_{-a}$ &
$1.70^{+0.51}_{-a}$ \\
& $\xi$ & erg cm s$^{-1}$ &
$1195^{+217}_{-636}$ &
$1214 \pm 409$ & 
$1100^{+202}_{-98}$ & 
$1094 \pm 547$ & 
$1088 \pm 558$ &
$1573^a$ \\
& $A_{\text{Fe}}$ &&
$0.41^{+1.41}_{-a}$ &
$0.66 \pm 0.47$ &
$0.1^{+0.5}_{-a}$ &
$3.09^a$ &
$0.1^{+2.0}_{-a}$ &
$1.01^{+14.46}_{-a}$ \\
$\chi^2 $ &&& 
0.60 & 
0.76 & 
0.75 & 
1.31 &
0.87 & 
0.99 \\
Absorbed ${F}_{X_{0.3-10.0\text{ keV}}}$ && $10^{-13}$ erg s$^{-1}$ cm$^{-2}$ &
$4.5^{+8.7}_{-3.8}$ &  
$2.7^{+2.6}_{-1.6}$ & 
$3.2^{+2.4}_{-2.1}$ &
$4.1^{+9.4}_{-3.8}$  &
$5.7^{+7.4}_{-4.3}$ &  
$1.4^b$\\
Unabsorbed ${F}_{X_{0.3-10.0\text{ keV}}}$ && $10^{-13}$ erg s$^{-1}$ cm$^{-2}$ &
$12.0^{+18.9}_{-9.5}$ &
$6.4^{+5.7}_{-3.8}$ & 
$8.5^{+5.7}_{-5.1}$ & 
$14.7^{+22.9}_{-10.7}$ & 
$12.1^{+11.9}_{-8.4}$ &
$1.4^b$  \\
\hline\hline
\end{tabular}
\begin{tablenotes}
    \small
    \item $^a$These parameters are not well constrained and errors are not reported.
    \item $^b$These values were found with \texttt{calc\_energy\_flux}. 
    \item $^*$Due to the lower total number of counts in 2011 relative to the other observations, the spectrum of ULX-2 at this epoch was grouped by 10 counts.
    \end{tablenotes}
\end{threeparttable}}
\end{center}
\end{specialtable}  
\begin{paracol}{2}
\switchcolumn

\newpage

\end{paracol}
\begin{specialtable}[H]	
\caption{\emph{XMM-Newton} ULX-1 and ULX-2 spectral models. \label{tab:xmm_spectra}}
\widefigure
\begin{center}
\scalebox{0.73}{
\begin{threeparttable}
\begin{tabular}{c c c | c c c | c c}
\hline\hline
\textbf{Component} & \textbf{Parameter} & \textbf{Units} & \multicolumn{3}{c|}{\textbf{ULX-1}} & \multicolumn{2}{c}{\textbf{ULX-2}} \\
&&& 2002 & 2011 & 2017 & 2011 & 2017  \\
\hline\hline
 $\text{TB}_{\text{ABS}} \cdot \text{BREMSS}$ \\
\hline\hline
$\text{TB}_{\text{ABS}}$ & $N_{H}$ & $10^{22}$ cm$^{-2}$ & 
$0.71 \pm 0.07$ & 
$0.47 \pm 0.02$ & 
$0.67 \pm 0.02$  & 
$0.194 \pm 0.004$  & 
$0.178 \pm 0.004$ \\
$\text{BREMSS}$  & $kT$ & keV & 
$4.0 \pm 0.7$ & 
$5.9 \pm 0.3$ & 
$6.6 \pm 0.3$  & 
$3.3 \pm 0.06$  & 
$4.6 \pm 0.1$ \\
$\chi^2$ &&& 
0.61 & 
0.89 & 
0.94 & 
1.37 & 
1.08 \\
Absorbed ${F}_{X_{0.3-10.0\text{ keV}}}$ && $10^{-13}$ erg s$^{-1}$ cm$^{-2}$ &
$3.0 \pm 0.5$ &  
$4.3 \pm 0.1$ & 
$3.8 \pm 0.1$ &
$16.7 \pm 0.3$ &
$4.9 \pm 0.1$ \\
Unabsorbed ${F}_{X_{0.3-10.0\text{ keV}}}$ && $10^{-13}$ erg s$^{-1}$ cm$^{-2}$ &
$4.8^{+0.7}_{-0.6} $ & 
$5.8 \pm 0.2$ & 
$5.4 \pm 0.1$ & 
$21.5 \pm 0.3$ & 
$6.04 \pm 0.09$ \\
\hline\hline
$\text{TB}_{\text{ABS}} \cdot \text{BKNPOWER}$ \\
\hline\hline
$\text{TB}_{\text{ABS}}$ & $N_{H}$ & $10^{22}$ cm$^{-2}$ & 
$1.32 \pm 0.31$ & 
$0.36 \pm 0.05$ & 
$0.69 \pm 0.03$  & 
$0.248 \pm 0.006$  & 
$0.197 \pm 0.008$ \\
BKNPOWER & $\Gamma_1$ && 
$3.3 \pm 0.9$ & 
$1.1 \pm 0.2$ & 
$1.61 \pm 0.04$  & 
$1.90 \pm 0.02$  & 
$1.64 \pm 0.03$ \\
& $\Gamma_2$ && 
$2.2 \pm 0.2$ & 
$2.04 \pm 0.07$ & 
$2.6 \pm 0.1$  & 
$3.44 \pm 0.09 $  & 
$2.9 \pm 0.1$ \\
& BreakE & keV &
$1.9 \pm 0.3$ & 
$2.0 \pm 0.2$ & 
$4.3 \pm 0.2$  & 
$3.8 \pm 0.1$  & 
$3.5 \pm 0.2$ \\
$\chi^2$ &&& 
0.55 &
0.88 & 
0.93 & 
1.26 & 
0.96 \\
Absorbed ${F}_{X_{0.3-10.0\text{ keV}}}$ && $10^{-13}$ erg s$^{-1}$ cm$^{-2}$ &
$3.3^{+3.0}_{-1.8}$ &  
$4.3^{+0.9}_{-0.7}$ & 
$3.8^{+0.3}_{-0.2}$ &
$16.6 \pm 0.4$  &
$4.9 \pm 0.2$ \\
Unabsorbed ${F}_{X_{0.3-10.0\text{ keV}}}$ && $10^{-13}$ erg s$^{-1}$ cm$^{-2}$ &
$21.4^{+19.5}_{-10.4}$ & 
$5.3 \pm 0.1$ & 
$5.6 \pm 0.3$  & 
$24.1 \pm 0.5$  & 
$6.2 \pm 0.2$ \\
\hline\hline
$\text{TB}_{\text{ABS}} \cdot \text{DISKPBB}$ \\
\hline\hline
$\text{TB}_{\text{ABS}}$ & $N_{H}$ & $10^{22}$ cm$^{-2}$ & 
$0.86^a$ &
$0.45 \pm 0.02$ &
$0.61 \pm 0.03$ &
$0.209 \pm 0.005$ &
$0.164 \pm 0.007$ \\
DISKPBB & $T_{\text{in}}$ & keV &
$2.4^a$ &
$2.1^{+0.2}_{-0.1}$ &
$2.2 \pm 0.1$ &
$1.56 \pm 0.02$ &
$1.72 \pm 0.02$ \\
& $p$ &&
$0.5^a$&
$0.587 \pm 0.005$ &
$0.61 \pm 0.01$ &
$0.552 \pm 0.003$ &
$0.592 \pm 0.005$ \\
$\chi^2 $ &&& 
0.59 & 
0.91 & 
0.88 & 
1.21 &
0.95 \\
Absorbed ${F}_{X_{0.3-10.0\text{ keV}}}$ && $10^{-13}$ erg s$^{-1}$ cm$^{-2}$ &
$3.1/3.7/3.6^*$ &  
$4.2^{+1.8}_{-1.6}$ & 
$3.7^{+1.1}_{-0.9}$ &
$16.4^{+1.3}_{-1.2}$  &
$4.8 \pm 0.4$ \\
Unabsorbed ${F}_{X_{0.3-10.0\text{ keV}}}$ && $10^{-13}$ erg s$^{-1}$ cm$^{-2}$ &
$3.1/3.7/3.6^*$ &
$5.7^{+2.4}_{-2.1}$ &
$5.0^{+1.5}_{-1.3}$ &
$22.0^{+1.8}_{-1.6} $ &
$5.8 \pm 0.5$ \\
\hline\hline
$\text{TB}_{\text{ABS}} \cdot (\text{DISKPBB}+\text{COMPTT})$ \\
\hline\hline
$\text{TB}_{\text{ABS}}$ & $N_{H}$ & $10^{22}$ cm$^{-2}$ & 
$0.91^{+0.11}_{-0.12}$ &
$0.48^a$ &
$0.63 \pm 0.01$ &
$0.22^a$ &
$0.26 \pm 0.03$ \\
DISKPBB & $T_{\text{in}}$ & keV & 
$1.59^a$ &
$2.13^a$ &
$2.1^a$ &
$1.33^a$ &
$1.70 \pm 0.03$ \\
& $p$ && 
$0.5$ (f) &
$0.59$ (f) &
$0.61$ (f) &
$0.55$ (f) &
$0.59$ (f) \\
COMPTT & $T0$ & keV &
$0.07^a$ &
$0.06^a$ &
$0.1^a$ & 
$0.01^a$ &
$0.07 \pm 0.01$ \\
& $kT_{e}$ & keV &
$23.4^a$ &
$44.3^a$ &
$51.1^a$ &
$48.5^a$ &
$42.9^a$ \\
& $\tau_p$ &&
$1.04^a$ &
$0.65^a$ &
$1.00^a$ &
$0.81^a$ &
$0.018 \pm 0.017$ \\
$\chi^2 $ &&& 
0.61 & 
0.91 & 
0.90 &
1.41 & 
0.92 \\
Absorbed ${F}_{X_{0.3-10.0\text{ keV}}}$ && $10^{-13}$ erg s$^{-1}$ cm$^{-2}$ &
$2.3/3.4/4.0^*$ &
$4.5/4.5/4.5^*$ &
$3.8/4.1/3.6^*$ &
$16.5/16.1/16.0^*$ &
$4.8/5.0/5.0^*$\\
Unabsorbed ${F}_{X_{0.3-10.0\text{ keV}}}$ && $10^{-13}$ erg s$^{-1}$ cm$^{-2}$ &
$2.3/3.4/4.0^*$ &
$4.5/4.5/4.5^*$ &
$3.8/4.1/3.6^*$ &
$16.5/16.1/16.0^*$ &
$4.8/5.0/5.0^*$ \\
\hline\hline
$\text{TB}_{\text{ABS}} \cdot \text{DISKIR}$ \\
\hline\hline
$\text{TB}_{\text{ABS}}$ & $N_{H}$ & $10^{22}$ cm$^{-2}$ &
$0.92^a$ &
$0.37^a$ &
$0.53 \pm 0.10$ &
$0.175^a$ &
$0.17^a$ \\
DISKIR & $kT_{\text{disk}}$ & keV & 
$0.42^a$ &
$1.0^a$ &
$1.0^a$ &
$0.89^a$ &
$1.16^a$ \\
& $\Gamma$ &&
$5.0^a$ &
$2.1^a$ &
$2.15 \pm 0.31$ &
$5.0^a$ &
$4.09^a$ \\
& $L_c/L_d$ && 
$0.90^a$ &
$0.78^a$ &
$1.24^a$ &
$0.39^a$ &
$0.36^a$ \\
& $kT_{e}$ & keV &
$37.8^a$ &
$5.1^a$ &
$6.5 \pm 4.5$ &
$5.0^a$ &
$407^a$ \\
& $f_{\text{in}}$ &&
$0.1$ (f) & 
$0.1$ (f) & 
$0.1$ (f) & 
$0.1$ (f) &
$0.1$ (f) \\
& $f_{\text{out}}$ && 
$0.1^a$ &
$0.02^a$ &
$0.02^{+0.05}_{-a}$ &
$0.10^a$ &
$0.1^a$ \\
& $r_{\text{irr}}$ & $r_{\text{in}}$ &
$1.002^a$ &
$1.02$ &
$1.2 \pm 0.1$ &
$1.01^a$ &
$1.02^a$ \\
& $\log r_{\text{out}}$ & $\log r_{\text{in}}$ & 
$5$ (f) &
$5$ (f) &
$5$ (f) &
$5$ (f) &
$5$ (f) \\
$\chi^2 $ &&& 
0.57 & 
0.86 & 
0.92 &
1.50 & 
0.97 \\
Absorbed ${F}_{X_{0.3-10.0\text{ keV}}}$ && $10^{-13}$ erg s$^{-1}$ cm$^{-2}$ &
$3.1/3.7/3.7^*$ &
$4.3/4.4/4.5^*$ &
$3.8/4.0/3.7^*$ &
$16.6/16.0/16.1^* $ &
$4.9/5.1/5.0^*$  \\
Unabsorbed ${F}_{X_{0.3-10.0\text{ keV}}}$ && $10^{-13}$ erg s$^{-1}$ cm$^{-2}$ &
$3.1/3.7/3.7^*$ &
$4.3/4.4/4.5^*$ &
$3.8/4.0/3.7^*$ &
$16.6/16.0/16.1^* $ &
$4.9/5.1/5.0^*$  \\
\hline\hline
$\text{TB}_{\text{ABS}} \cdot \text{KDBLUR2} \cdot \text{REFLIONX}$ \\
\hline\hline
$\text{TB}_{\text{ABS}}$ & $N_{H}$ & $10^{22}$ cm$^{-2}$ &
$1.52^{+0.26}_{-0.31}$ &
$2.27 \pm 0.32$ &
$1.09^{+7.07}_{-0.23} $ &
$2.12 \pm 0.07$ &
$1.57 \pm 0.05$ \\
KDBLUR2 & $R_{\text{in}}$ & $R_G$ &
$8.2^{+2.9}_{-a}$ &
$1.24^{+0.13}_{-a}$ &
$4.43^{+0.75}_{-a}$ &
$1.38 \pm 4.2\mathrm{e}{-6}$ &
$1.38^{+0.23}_{-a}$ \\
& $R_{\text{out}}$ & $R_G$ &
$400$ (f) &
$400$ (f) &
$400$ (f) &
$400$ (f) &
$400$ (f) \\
& $i$ & deg & 
$38.4^{+12.3}_{-a}$ &
$61.2 \pm 5.0$ &
$61.3^{+26.6}_{-a} $ &
$35.3 \pm 2.4$ &
$30.8 \pm 2.3$  \\
& $q_{\text{in}}$ &&
$10.0^{+a}_{-18.4}$ &
$7.6 \pm 1.2$ &
$8.9^{+a}_{-2.6}$ &
$7.9 \pm 0.4$ &
$6.6 \pm 0.2$ \\
& $q_{\text{out}}$ && 
$3.0$ (f) & 
$3.0$ (f) &
$3.0$ (f) &
$3.0$ (f) &
$3.0$ (f) \\
& $R_{\text{break}}$ & $R_G$ &
$20.0$ (f) &
$20.0$ (f) &
$20.0$ (f) &
$20.0$ (f) &
$20.0$ (f) \\
REFLIONX & $\Gamma$ &&
$2.3^{+0.36}_{-0.15}$ &
$2.03 \pm 0.07$ &
$1.90 \pm 0.06$ &
$2.19 \pm 0.02$ &
$1.8^a$ \\
& $\xi$ & erg cm s$^{-1}$ &
$1229^{+a}_{-1047}$ &
$1365 \pm 257$ &
$1141^{+81}_{-12}$ &
$1492 \pm 123$ &
$1143 \pm 37$ \\
& $A_{\text{Fe}}$ &&
$0.90^{+0.88}_{-0.47}$ &
$0.11^{+0.20}_{-a}$ &
$0.29^{+0.08}_{-0.10}$ &
$1.08 \pm 0.21$ &
$1.06^{+2.20}_{-0.08}$ \\
$\chi^2 $ &&& 
0.66 & 
0.80 & 
1.27 &
1.62 & 
1.21 \\
Absorbed ${F}_{X_{0.3-10.0\text{ keV}}}$ && $10^{-13}$ erg s$^{-1}$ cm$^{-2}$ &
$3.7^{+3.0}_{-2.5}$ &  
$4.4^{+1.6}_{-1.3}$ & 
$4.2^{+2.6}_{-2.3}$ &
$16.6^{+2.6}_{-2.5}$  &
$4.9 \pm 0.5$ \\
Unabsorbed ${F}_{X_{0.3-10.0\text{ keV}}}$ && $10^{-13}$ erg s$^{-1}$ cm$^{-2}$ &
$9.0^{+6.8}_{-6.2}$ &
$10.8^{+3.9}_{-3.1}$ &
$9.9^{+6.0}_{-5.5}$ &
$37.8 \pm 5.5$ &
$8.8 \pm 1.1$ \\
\hline\hline
\end{tabular}
    \begin{tablenotes}
      \small
      \item $^a$These parameters are not well constrained and errors are not reported.
      \item $^*$Sample flux does not converge, values found via \texttt{calc\_energy\_flux} for pn/mos1/mos2
    \end{tablenotes}
\end{threeparttable}}
\end{center}
\end{specialtable} 
\begin{paracol}{2}
\switchcolumn

\section{Discussion}
\subsection{ULX-1}
ULX-1 has been visible since at least the 1990s, when it was referred to as ``XNorth''  in observations with \emph{ROSAT} \citep{1998ApJ...493..431H}. However, the source still remains poorly characterized. It is difficult to distinguish between the various models, as shown comprehensively in Table~\ref{tab:chandra_spectra} and demonstratively for the 2003 epoch in Figure~\ref{fig:ULX1_2003_chandra}. The fits appear to be statistically similar, such that there is no single best-fit model that stands out. This makes it difficult to delineate the radiative mechanism giving rise to the observed emission. 

\begin{figure}[H]
\centering
\includegraphics[width=14cm]{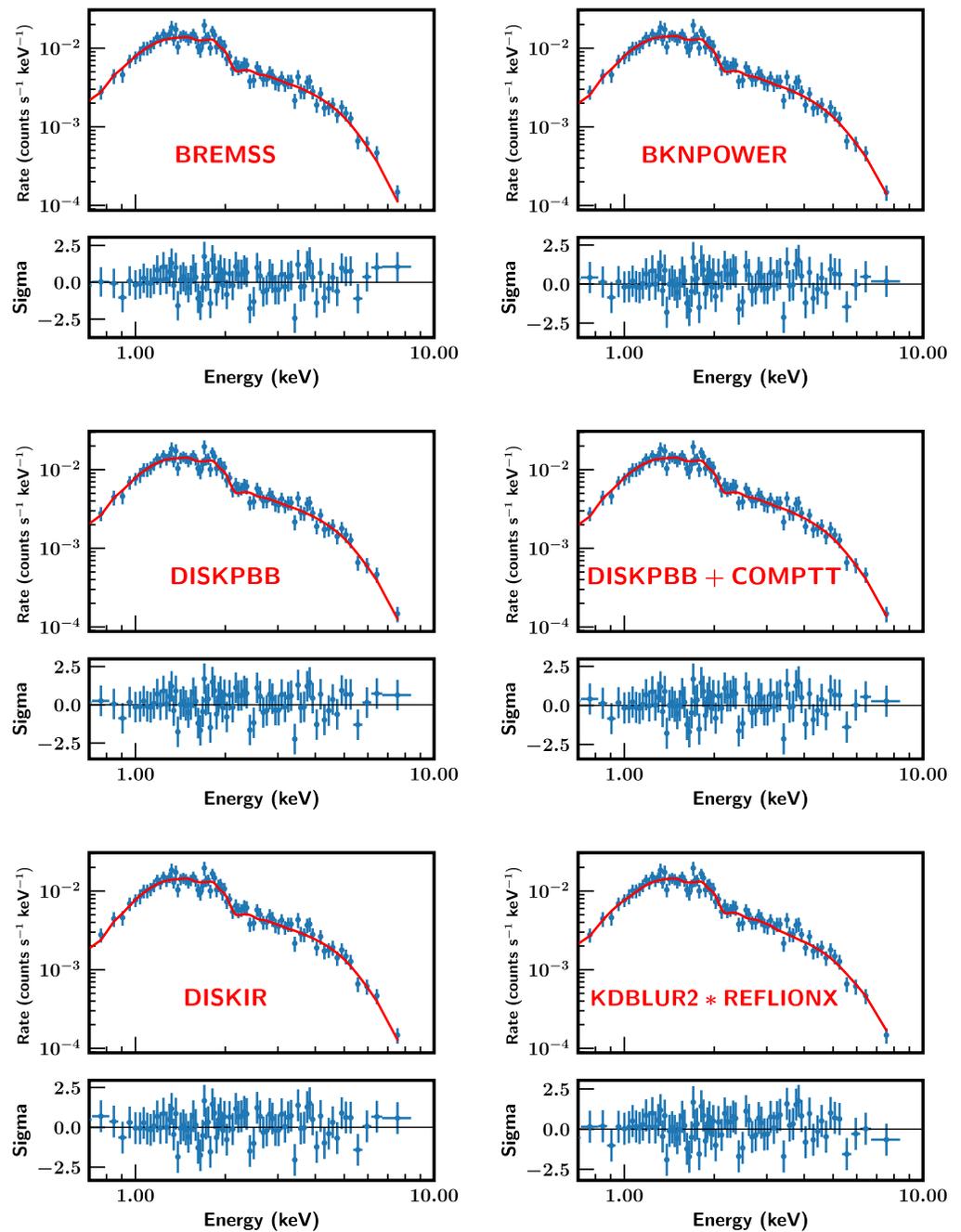}
\caption{Fits with thermal, power-law, and accretion disk models of ULX-1's spectrum as observed by \emph{Chandra} in 2003. Each model includes a thermal absorption component that was left free to vary. The spectrum is grouped by 30 counts. Best-fit parameters are given explicitly in Table~\ref{tab:chandra_spectra}. The residuals corresponding to each model indicate highly similar fits. \label{fig:ULX1_2003_chandra}}
\end{figure}

The spectral fitting with \emph{Chandra} does indicate a generally consistent high column density over all models and epochs, with $N_H \sim 10^{22} \textrm{ cm}^{-2}$, in excess of the Galactic column of 0.07 $\times 10^{22}$ cm$^{-2}$ towards SN 1986J \citep{dl90,kalberlaetal05, h14pi}. However, NGC 891 is viewed almost edge-on, and therefore one also needs to take into account the column density through the disk of that galaxy. This depends on the line of sight but could have a value of $\log{N_H} =20.7$ \citep{dasetal20}, or could be much higher and almost comparable in value to the observed column density \citep{aokietal91}. This also depends on whether the ULX is physically located near the SN or just happens to be viewed along a similar line-of-sight. As shown in Table~\ref{tab:chandra_spectra}, the ULX fits reveal a different column density for each ULX, which could mean that they are located at different depths in the galaxy, or that they have differing amounts of material surrounding them, or a combination of the two. 

The column densities found for ULX-1 via fitting \emph{Chandra} and \emph{XMM-Newton} spectra (Tables~\ref{tab:chandra_spectra} and \ref{tab:xmm_spectra}) are generally consistent in 2002, with $N_H \sim 10^{22} \text{ cm}^{-2}$. In 2011, the \emph{XMM} $N_H$ decreases to almost half its 2002 value. In 2017, \emph{XMM} gives slightly larger values for $N_H$ in comparison to 2011, but still less than the 2016 value found from the \emph{Chandra} spectra.  We do not fit the \emph{Chandra} spectra of ULX-1 in 2011 due to the short exposure time of the observation. We find no decrease in $N_H$ in the \emph{Chandra} results between 2000 and 2016. This discrepancy in $N_H$ between the \emph{Chandra} and \emph{XMM} data could possibly be due to the proximity of ULX-2 to ULX-1, given the different PSFs of the two telescopes. ULX-2 appeared in 2011 and persisted through 2017 at least. During those years, the large \emph{XMM} PSF suggests that the flux from ULX-1 may have been contaminated by that from ULX-2, although we have tried to isolate the emission as best as possible. Fitting the data results in a column density which lies somewhere between that of ULX-1 and ULX-2.

The hot disk model (\texttt{diskpbb}) returns very high best-fit input temperatures. The value of $p$ is often, but not always, close to 0.5, suggesting the possibility for advective flow in a disk. The temperatures are lowered by adding a Comptonization component, but are still $> 1$ keV and sometimes $> 2$ keV. Thus the disk remains `hot'. The addition of the Comptonization component does not do much to improve the fit. The intrinsic absorption in a model where the inner disk is irradiated by Comptonized photons can be lower than the previous models, but both the disk and corona temperatures are high. None of the fits produce cool disks; namely, disk temperatures generally hover around 1 keV for both the \emph{Chandra} and \emph{XMM} fits. This model results in cooler disk temperatures, but a high corona temperature. The blurred reflection model returns a higher intrinsic absorption with a generally high emissivity index $q_{in}$, implying disks that are dominated by the emission from the inner disk which falls off steeply with radius. The disk is generally found to have a high inclination angle, although the error bars can be large. The inner edge of the disk extends closer to the last stable orbit in the last Chandra observation, but not in the first two.  The ionization parameter in all cases is extremely large, and the Fe abundance sub-solar. For these models with multiple parameters, the \texttt{sample\_flux} command at times fails to converge. We have calculated the flux using \texttt{calc\_energy\_flux}, which unfortunately does not allow calculation of the error bars. 

The \emph{Chandra} models show some spectral evolution of the source from 2003 to 2016. The column density in all models is high. The unabsorbed flux
(0.3-10.0 keV band) over all epochs has a value generally around $5-10 \times 10^{-13}$ erg s$^{-1}$ cm$^{-2}$. Using the \emph{ROSAT} PSPC and HRI count rates in PIMMS, and assuming a power-law model with the same photon index found from the \emph{Chandra} spectral fit, we calculate the unabsorbed fluxes from the August 1991, July 1993, and January 1995 data sets to be $5.6 \times 10^{-13} \text{ erg} \text{ s}^{-1} \text{ cm}^{-2}$, $5.3 \times 10^{-13} \text{ erg} \text{ s}^{-1} \text{ cm}^{-2} $, and $6.8 \times 10^{-13} \text{ erg} \text{ s}^{-1} \text{ cm}^{-2} $ respectively. These are consistent with the \emph{Chandra} and \emph{XMM-Newton} values.

The light curve shows a possible slight decrease in flux over time, particularly from 2000 to 2003, although the error bars are large enough that a steady flux over the entire time period is also valid. There does not seem to have been significant flux evolution over almost 30 years. At a distance of 9 Mpc, this flux corresponds to an isotropic luminosity of $L_X \sim 5-10 \times 10^{39}$ erg s$^{-1}$, consistent with the luminosities of other ULXs \citep{Kovlakas2020}. \citet{swartzetal11}  analyzed the \emph{Chandra} 2000 observation of ULX-1 assuming a power-law model. At the distance of 9 Mpc used in our work, their result corresponds to a luminosity of $8.4 \times 10^{39}$ erg s$^{-1}$, consistent with our derived values. Our flux values are also in agreement with those quoted by \citet{dageetal21} for the 2000 observation. The long-term stability of the light curve suggests that the source is not a highly variable object over these timescales (Figure~\ref{flux_plots}).

\begin{figure}[H]
\centering
\includegraphics[width=10.5 cm]{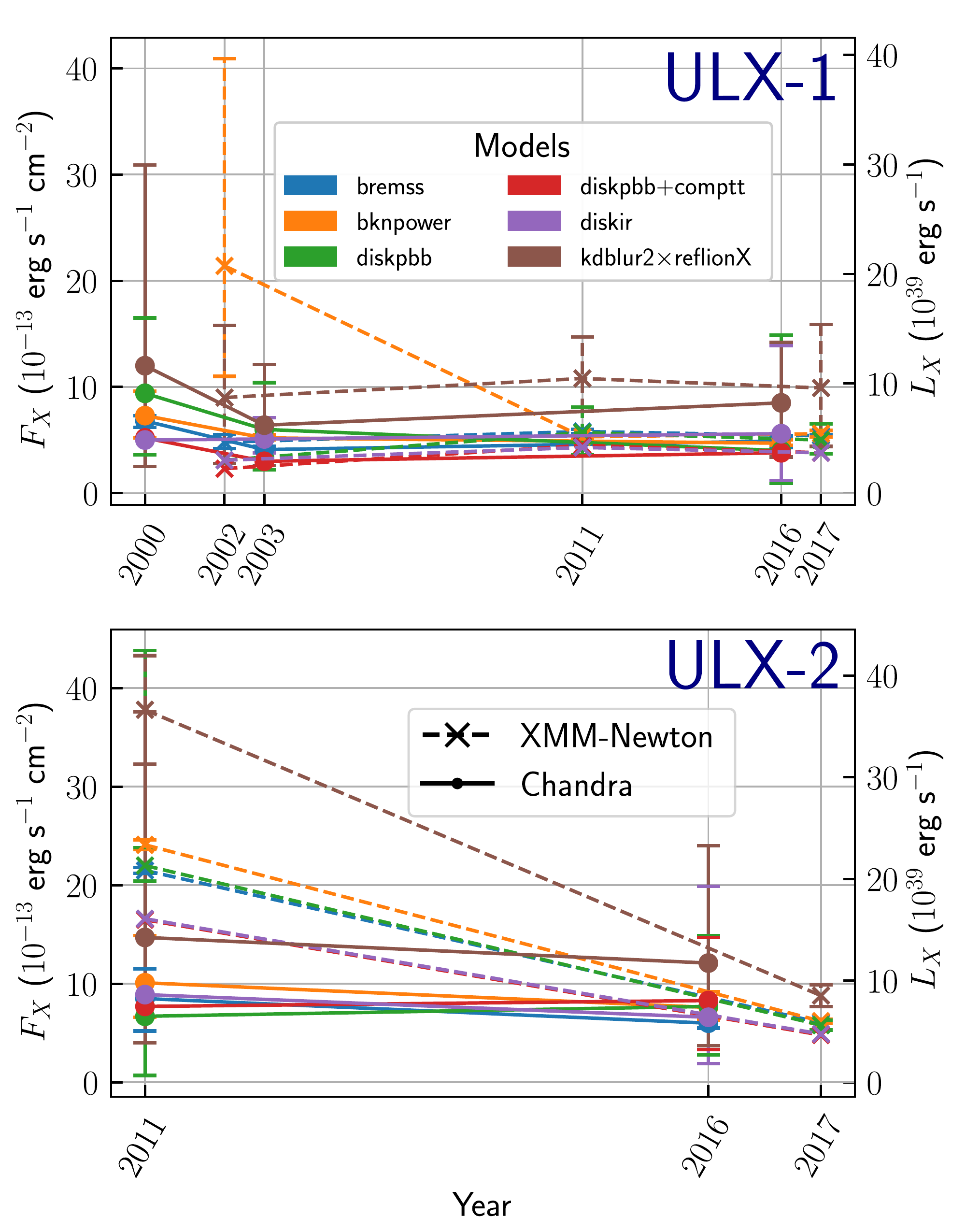}
\caption{The unabsorbed model light curves for ULX-1 and ULX-2 using \emph{Chandra} and \emph{XMM-Newton} spectra. The derived flux for any particular model for ULX-1 is consistent in time within error bars, suggesting long term stability. Due to the wide spread in unabsorbed flux in 2011, because of low statistics, it is not possible to determine whether ULX-2 reduced in luminosity from 2011 to 2017. \label{flux_plots}}
\end{figure}

\subsection{ULX-2}
ULX-2 is not visible in either the 2000 or 2003 \emph{Chandra} spectra. We calculate 90\% confidence upper limits for non-detections, using a 4'' radius source region, in the 2000 and 2003 data. Using the \texttt{srcflux} command in CIAO and assuming a simple redshifted power-law with $N_H = 0.2 \times 10^{22}$ cm$^{-2}$, we find that in the \emph{Chandra} 2000 dataset, the absorbed and unabsorbed upper limits are $4.0$ and $4.9 \times 10^{-15} \text{ erg} \text{ s}^{-1} \text{ cm}^{-2}$, respectively. This is almost a factor of 200 less than the detected source flux in 2011. In 2003, the upper limits are similarly constraining, with absorbed and unabsorbed flux values near $1.3$ and $1.4 \times 10^{-15} \text{ erg} \text{ s}^{-1} \text{ cm}^{-2}$, respectively. Clearly, if there was a source present at that position in these datasets, its luminosity was at least two orders of magnitude lower than when it was detected.
HK12 first detected ULX-2 with \emph{XMM-Newton} in 2011 and fit the spectrum. Due to \emph{XMM-Newton's} limited spatial resolution, it is possible that emission from ULX-2 could be contaminated by emission from the nearby SN 1986J and ULX-1. The much better spatial resolution of the \emph{Chandra} satellite reduces or eliminates the contamination.  The spectrum and fits using the various models for the 2016 epoch are shown in Figure~\ref{fig:ULX2_2016_chandra}. As before with ULX-1, all the fits are found to be similar, both to the eye and statistically. Excess emission near 1.8 keV is difficult to fit with any of the spectral models. 

\begin{figure}[H]
\centering
\includegraphics[width=14cm]{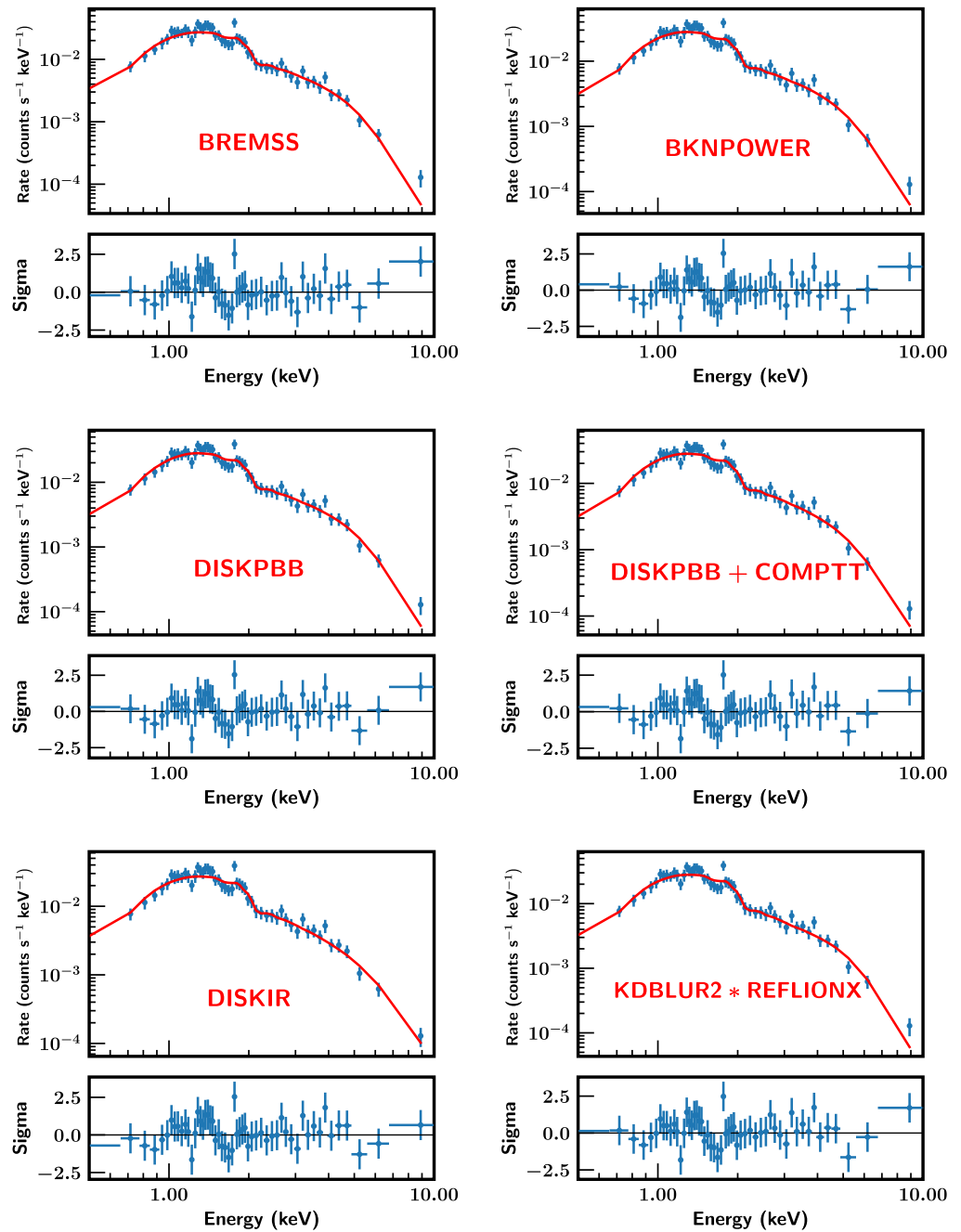}
\caption{The same as Figure~\ref{fig:ULX1_2003_chandra}, except for ULX-2's spectrum as observed by \emph{Chandra} in 2016. \label{fig:ULX2_2016_chandra}}
\end{figure}

The absorption deduced for the two simpler models is comparable to that seen by HK12, but overall the similarities between the values are limited. The temperature returned by the thermal bremsstrahlung model in the \emph{Chandra} fit is much higher than that found by HK12, although their value is comparable to our \emph{XMM-Newton} fit. The broken power-law model is also more consistent with the \emph{XMM-Newton} fit than with \emph{Chandra}. The story is the same with the hot disk model, where the disk temperature is found to be higher with \emph{Chandra}. Adding a Comptonized component does not decrease the temperature, although the \emph{XMM-Newton} model is closer to the result of HK12. HK12 find a lower level of absorption in the \texttt{diskir} model, although our derived temperatures for both the disk and the corona are higher. The blurred reflection models in both the \emph{Chandra} and \emph{XMM} cases approach the results of HK12 in many of the parameters. Overall, even given the higher flux and the larger number of counts compared to ULX-1, it is difficult to select any model as being superior to the others. Similar to ULX-1, the temperatures returned by \emph{Chandra} generally seem higher than that obtained with \emph{XMM}. Our 2011 \emph{XMM-Newton} values are in agreement with those derived by HK12, and the 2017 values are consistent with those obtained by \citet{hketal18} using a p-free blackbody disk model. 

ULX-2 was observed in 2011 and 2016 with \emph{Chandra}. The unabsorbed flux in the various models is about $6-12 \times 10^{-13}$ erg s$^{-1}$ cm$^{-2}$ in the 0.3-10.0 keV band. \citet{dageetal21} quote a flux of $8.8 \pm 0.2$ erg s$^{-1}$ cm$^{-2}$ using an absorbed power-law model, and $6.7 \pm 2.0$ erg s$^{-1}$ cm$^{-2}$ using an absorbed multiblackbody disk model for the 2011 observation, in agreement with our results. The flux is remarkably constant over most models in 2016 except for the last reflection model, which is higher than all the others, but still within a factor of 2. The variations are larger in 2011, primarily in the flux calculated with the \texttt{kdblur2*reflionX} models, which is not surprising given the low spectral counts. The flux appears to be 20-50\% higher than that of ULX-1, which is expected from the count rates. Again, while the best-fit values may reflect a decrease over 5 years, especially as seen by \emph{XMM}, within the error bars the ULX may have had a consistent luminosity over the same time period (Figure~\ref{flux_plots}).  The derived column density is around $0.20 \times 10^{22}$ cm$^{-2}$  with some variations. This value is lower than that found for ULX-1 by a factor of a few. This could either mean that the two are at different depths in the galaxy and not spatially close to each other, or it could mean that there is a differing amount of material around the two ULXs. The column density for ULX-2 is lower than any of the column densities calculated for the other ULXs.

\subsection{ULX-3}
Similar to ULX-2, ULX-3 is also not detected in 2000 or 2003 \emph{Chandra} spectra. We calculate 90\% confidence upper limits for non-detections in 2000 and 2003, using a redshifted power-law with $N_H = 2.0 \times 10^{22}$ cm$^{-2}$ as found from spectral fitting in 2016. In 2000, the absorbed and unabsorbed upper limits are $2.9$ and $3.5 \times 10^{-15} \text{ erg} \text{ s}^{-1} \text{ cm}^{-2}$, respectively. In 2003, the absorbed and unabsorbed flux values are $1.1$ and $1.2 \times 10^{-15} \text{ erg} \text{ s}^{-1} \text{ cm}^{-2}$, respectively. Thus the flux at the ULX-3 position at previous epochs is at least two orders of magnitude lower than the detection value. ULX-3 is first detected in the \emph{Chandra} Nov 2016
observation. Fitting the \emph{Chandra} spectrum of ULX-3 with the same
set of models as used for the other sources results in an unabsorbed
(0.3-10.0 keV) flux around $2 \times 10^{-13} \text{ erg} \text{ s}^{-1}
\text{ cm}^{-2}$. ULX-3 is the faintest of the three ULXs. At a distance of 9 Mpc, $L_X = 2 \times 10^{39} \text{ erg} \text{ s}^{-1}$, which places the source in the lower luminosity range of detected ULXs. The absorption column on the other hand exceeds that found for ULX-1 or ULX-2 by more than a factor of two. The source temperature in the thermal model is extremely high ($kT > 20$ keV). The reduced $\chi^2$ is roughly equivalent between the
thermal bremsstrahlung, power-law, and p-free multicolor disk
blackbody models, but improves when more complicated disk models are
considered. The fits for the Nov 2016 observation are shown in the last column of Table~\ref{tab:chandra_spectra}, to enable comparison with the \emph{Chandra} spectra of the other sources, and the first column of Table~\ref{tab:ulx3_spectra}, to demonstrate the source's spectral evolution over time. Results for a redshifted power-law are shown in Table~\ref{tab:ulx3_spectra}. 

We simultaneously fit the EPIC-pn, MOS1, and MOS2 spectra, collected a couple months later. These results are shown in Table~\ref{tab:ulx3_spectra}. We primarily consider the bremsstrahlung and power-law models, since the parameters can be well constrained, and error bars on all the parameters can be obtained. The flux and column density around the source are both found to decrease by a factor of seven from Nov 2016 to Jan 2017. At this time the source no longer qualifies as  `ultraluminous'. This lower luminosity value is more consistent with other high energy sources such as X-ray binaries ($L_X > 10^{37} \textrm{ erg } \textrm{s}^{-1}$) \citep{Fabbiano2016}. 

\clearpage

\end{paracol}

\begin{specialtable}[H]	
\caption{2016 \emph{Chandra} and 2017 \emph{XMM-Newton} ULX-3 spectral models. \label{tab:ulx3_spectra}}
\widefigure
\begin{center}
\scalebox{0.73}{
\begin{threeparttable}
\begin{tabular}{c c c | c | c | c | c | c | c}
\hline\hline
\textbf{Component} & \textbf{Parameter} & \textbf{Units} &  \textbf{Nov 14 2016} & \textbf{Jan 27 2017} & \textbf{Jan 29 2017} & \textbf{Feb 19 2017} & \textbf{Feb 23 2017} & \textbf{Feb 25 2017}  \\
\hline\hline
$\text{TB}_{\text{ABS}} \cdot \text{BREMSS}$ \\
\hline\hline
$\text{TB}_{\text{ABS}}$ & $N_{H}$ & $10^{22}$ cm$^{-2}$ & 
$2.2 \pm 0.2$& 
$0.30 \pm 0.15$ & 
$0.24^{+0.09}_{-0.07}$  & 
$0.20 \pm 0.07$  & 
$0.29 \pm 0.12$ & 
$0.22 \pm 0.08$ \\
$\text{BREMSS}$  & $kT$ & keV & 
$24.1^{+a}_{-16.5}$ & 
$38.5^{+82.4}_{-a}$ & 
$12.7^{+13.5}_{-5.1}$  & 
$9.49 \pm 4.44$  & 
$4.68 \pm 2.05$ & 
$7.08 \pm 3.01$ \\
$\chi^2$ &&& 
0.67 & 
0.89 & 
0.54 & 
0.67 & 
0.49 & 
0.92 \\
Absorbed ${F}_{X_{0.3-10.0\text{ keV}}}$ && $10^{-13}$ erg s$^{-1}$ cm$^{-2}$ &
$1.4 \pm 0.1$ &  
$0.31^*$ & 
$0.36^{+0.04}_{-0.05}$ &
$0.27^{+0.05}_{-0.06}$  &
$0.23^{+0.07}_{-0.09}$ &  
$0.21 \pm 0.05$\\
Unabsorbed ${F}_{X_{0.3-10.0\text{ keV}}}$ && $10^{-13}$ erg s$^{-1}$ cm$^{-2}$ &
$2.1 \pm 0.1$ &
$0.31^*$ & 
$0.42^{+0.05}_{-0.06} $  & 
$0.32 \pm 0.06$  & 
$0.30^{+0.08}_{-0.10}$ & 
$0.26^{+0.05}_{-0.06}$ \\
\hline\hline
$\text{TB}_{\text{ABS}} \cdot \text{ZPOWERLW}$ \\
\hline\hline
$\text{TB}_{\text{ABS}}$ & $N_{H}$ & $10^{22}$ cm$^{-2}$ & 
$2.2^{+1.0}_{-0.7}$ &
$0.29 \pm 0.20 $ & 
$0.28 \pm 0.10 $  & 
$0.26 \pm 0.11$ & 
$0.40 \pm 0.19$ & 
$0.28 \pm 0.11$ \\
ZPOWERLW & $\Gamma$ && 
$1.4 \pm 0.4$ & 
$1.28 \pm 0.25$ & 
$1.52 \pm 0.18$ & 
$1.63 \pm 0.23$ & 
$1.93 \pm 0.32$ & 
$1.71 \pm 0.22$ \\
$\chi^2 $ &&& 
0.68 &
0.90 & 
0.59 & 
0.68 & 
0.53 &
0.97 \\
Absorbed ${F}_{X_{0.3-10.0\text{ keV}}}$ && $10^{-13}$ erg s$^{-1}$ cm$^{-2}$ &
$1.4^{+1.7}_{-0.9}$ &  
$0.32^{+0.18}_{-0.13}$ & 
$0.39^{+0.14}_{-0.11}$ &
$0.29^{+0.15}_{-0.09}$  &
$0.26^{+0.15}_{-0.11}$ &  
$0.24^{+0.10}_{-0.08} $\\
Unabsorbed ${F}_{X_{0.3-10.0\text{ keV}}}$ && $10^{-13}$ erg s$^{-1}$ cm$^{-2}$ &
$2.2^{+2.0}_{-1.2}$ &
$0.36^{+0.19}_{-0.14}$ & 
$0.47^{+0.14}_{-0.12}$ & 
$0.36^{+0.14}_{-0.11}$ & 
$0.40^{+0.17}_{-0.14}$ & 
$0.31^{+0.09}_{-0.08}$ \\
\hline\hline
$\text{TB}_{\text{ABS}} \cdot \text{DISKPBB}$ \\
\hline\hline
$\text{TB}_{\text{ABS}}$ & $N_{H}$ & $10^{22}$ cm$^{-2}$ & 
$1.45 \pm 0.81$ &
$0.09^a$ &
$0.08^{+0.13}_{-0.08}$ &
$0.18 \pm 0.10$ &
$0.07^a$ &
$0.07^{+0.16}_{-a}$ \\
DISKPBB & $T_{\text{in}}$ & keV &
$2.0 \pm 0.8$ &
$1.7^a$ &
$1.8 \pm 0.4$ &
$2.5^{+a}_{-1.0}$ &
$1.1^a$ &
$1.4 \pm 0.4$ \\
& $p$ &&
$1.0 \pm 0.8$ &
$1.0^a$ &
$0.82 \pm 0.21$ &
$0.61 \pm 0.02$ &
$1.0^a $ &
$0.83 \pm 0.26$ \\
$\chi^2 $ &&& 
0.65 & 
0.83 & 
0.45 & 
0.68 &
0.46 &  
0.89 \\
Absorbed ${F}_{X_{0.3-10.0\text{ keV}}}$ && $10^{-13}$ erg s$^{-1}$ cm$^{-2}$ &
$2.0^{+8.8}_{-1.8}$ &  
$0.26^*$ & 
$0.36^{+0.89}_{-0.28}$ &
$0.28^*$ &
$0.22^*$ & 
$0.21^{+0.62}_{-0.17}$\\
Unabsorbed ${F}_{X_{0.3-10.0\text{ keV}}}$ && $10^{-13}$ erg s$^{-1}$ cm$^{-2}$ &
$2.7^{+11.4}_{-2.5} $ &
$0.26^*$ &
$0.37^{+0.94}_{-0.28}$ &
$0.28^*$ &
$0.22^*$ &
$0.23^{+0.69}_{-0.19} $ \\
\hline\hline
$\text{TB}_{\text{ABS}} \cdot (\text{DISKPBB}+\text{COMPTT})$ \\
\hline\hline
$\text{TB}_{\text{ABS}}$ & $N_{H}$ & $10^{22}$ cm$^{-2}$ & 
$1.82^{+1.56}_{-1.79}$ &
$0.34^a$ &
$0.13^{+0.07}_{-a}$ &
$0.24^{+0.13}_{-a} $ &
$0.11^a$ &
$0.18^{+0.46}_{-a}$ \\
DISKPBB & $T_{\text{in}}$ & keV & 
$0.47 \pm 9.4\mathrm{e-}5 $ &
$1.1^a$ &
$1.5^a$ &
$0.73^a$ &
$1.04^a$ &
$1.35^a$ \\
& $p$ && 
1 (f) &
1 (f) &
$0.82$ (f) &
$0.61$ (f) &
1 (f) &
$0.83$ (f) \\
COMPTT & $T0$ & keV &
$0.35^{+4.86}_{-a}$ & 
$0.1^a$ & 
$0.05^a$ &
$0.1^a$ &
$0.08^a$ &
$0.06^a$ \\
& $kT_{e}$ & keV &
$50.7^a$ &
$59.0^a$ &
$41.5^a$ &
$31.5^a$ &
$26.6^a$ &
$38.8^a$ \\
& $\tau_p$ &&
$2.6 \pm 2.3$ &
$4.2^a$ &
$4.3^a$ &
$1.9^a$ &
$3.1^a$ &
$1.3^a$ \\
$\chi^2 $ &&& 
0.96 & 
0.93 & 
0.51 & 
0.82 &
0.54 & 
1.04 \\
Absorbed ${F}_{X_{0.3-10.0\text{ keV}}}$ && $10^{-13}$ erg s$^{-1}$ cm$^{-2}$ &
$0.6^{+1.6}_{-0.5}$ &  
$0.31^*$ & 
$0.34^{+0.36}_{-0.21}$ &
$0.38^*$ &
$0.25^*$ &
$0.27^*$ \\
Unabsorbed ${F}_{X_{0.3-10.0\text{ keV}}}$ && $10^{-13}$ erg s$^{-1}$ cm$^{-2}$ &
$0.7^{+1.8}_{-0.5}$ &
$0.31^*$ &
$0.50^{+0.51}_{-0.28}$ &
$0.38^*$ &
$0.25^*$ &
$0.27^*$ \\
\hline\hline
$\text{TB}_{\text{ABS}} \cdot \text{DISKIR}$ \\
\hline\hline
$\text{TB}_{\text{ABS}}$ & $N_{H}$ & $10^{22}$ cm$^{-2}$ &
$2.82 \pm 0.22$ &
$0.99^a$ &
$0.09^a$ &
$0.17^a$ &
$0.09^a$ &
$0.07^a$  \\
DISKIR & $kT_{\text{disk}}$ & keV & 
$0.17^{+0.12}_{-0.10}$ &
$1.0^a$ &
$0.92^a$ &
$0.90^a$ &
$0.68^a$ &
$0.94^a$ \\
& $\Gamma$ &&
$1.5 \pm 0.2$ &
$3.1^a$ &
$5.0^a $ &
$5.0^a$ &
$4.1^a$ &
$5.0^a$ \\
& $L_c/L_d$ && 
$10.0^a$ &
$0.2^a$ &
$2.0^a $ &
$1.3^a$ &
$2.2^a $ &
$1.2^a$ \\
& $kT_{e}$ & keV &
$142.2 \pm 33.9 $ &
$435.1^a$ &
$7.3^a$ &
$6.7^a$ &
$12.8^a$ &
$5.1^a$ \\
& $f_{\text{in}}$ &&
$0.15$ (f) & 
$0.1$ (f) & 
$0.1$ (f) & 
$0.1$ (f) & 
$0.1$ (f) &
$0.1$ (f) \\
& $f_{\text{out}}$ && 
$4.2\mathrm{e-}6^a$ &
$7.7\mathrm{e-}4^a$ &
$0.004^a$ &
$0.03^a$ &
$1.3\mathrm{e-}5^a $ &
$0.003^a$ \\
& $r_{\text{irr}}$ & $r_{\text{in}}$ &
$3.21^a$ &
$1.03^a$ &
$1.17^a$ &
$1.02^a$ &
$1.59^a$ &
$1.68^a$ \\
& $\log r_{\text{out}}$ & $\log r_{\text{in}}$ &
$5$ (f) & 
$5$ (f) & 
$5$ (f) & 
$5$ (f) & 
$5$ (f) &
$5$ (f) \\
$\chi^2 $ &&& 
0.995 & 
1.32 & 
0.50 & 
0.80 &
0.54 & 
1.02 \\
Absorbed ${F}_{X_{0.3-10.0\text{ keV}}}$ && $10^{-13}$ erg s$^{-1}$ cm$^{-2}$ &
$1.4^*$ &
$0.19^*$ &
$0.35^*$ &
$0.28^*$ &
$0.22^*$ &
$0.20^*$ \\
Unabsorbed ${F}_{X_{0.3-10.0\text{ keV}}}$ && $10^{-13}$ erg s$^{-1}$ cm$^{-2}$ &
$1.4^*$ &
$0.19^*$ &
$0.35^*$ &
$0.28^*$ &
$0.22^*$ &
$0.20^*$ \\
\hline\hline
$\text{TB}_{\text{ABS}} \cdot \text{KDBLUR2} \cdot \text{REFLIONX}$ \\
\hline\hline
$\text{TB}_{\text{ABS}}$ & $N_{H}$ & $10^{22}$ cm$^{-2}$ &
$2.48^{+1.92}_{-0.47} $ &
$1.02 \pm 0.13$ &
$2.24^{+0.93}_{-0.33} $ &
$1.37 \pm 0.22$ &
 &
$2.41^{+0.90}_{-0.51}$  \\
KDBLUR2 & $R_{\text{in}}$ & $R_G$ &
$1.24^{+8.17}_{-a} $ &
$2.60 \pm 0.68$ &
$1.78^{+0.72}_{-a} $ &
$2.09 \pm 0.12$ &
 &
$1.78^{+0.41}_{-0.45}$ \\
& $R_{\text{out}}$ & $R_G$ &
$400$ (f) &
$400$ (f) &
$400$ (f) &
$400$ (f) &
 &
$400$ (f) \\
& $i$ & deg & 
$90.0^{+a}_{-38.7} $ &
$85.4 \pm 1.1$ &
$29.4^{+9.2}_{-a} $ &
$45.8 \pm 5.5$ &
 &
$0^{+28}_{-a} $ \\
& $q_{\text{in}}$ &&
$9.5^{+a}_{-16.3} $ &
$9.4 \pm 0.3$ &
$6.6^{+3.2}_{-2.1} $ &
$10.0^{+a}_{-7.2} $ &
 &
$6.6^{+a}_{-1.0} $\\
& $q_{\text{out}}$ && 
$3.0$ (f) & 
$3.0$ (f) & 
$3.0$ (f) &
$3.0$ (f) &
 &
$3.0$  (f) \\
& $R_{\text{break}}$ & $R_G$ &
$20.0$ (f) & 
$20.0$ (f) &
$20.0$ (f) &
$20.0$ (f) &
 &
$20.0$ (f) \\
REFLIONX & $\Gamma$ &&
$1.70^{+0.51}_{-a} $ & 
$1.70^{+0.40}_{-0.19}$ &
$1.78^{+0.59}_{-a} $ &
$1.69 \pm 0.13$ &
 &
$1.60^{+0.67}_{-a}$ \\
& $\xi$ & erg cm s$^{-1}$ &
$1573^a$ &
$1004^{+325}_{-473}$& 
$1018^{+1317}_{-657}$ & 
$1060 \pm 124$ & 
 &
$1152^{208}_{-577} $ \\
& $A_{\text{Fe}}$ &&
$1.01^{+14.46}_{-a} $ &
$0.10^{+0.45}_{-a}$ &
$19.3^{+a}_{-14.6} $ &
$5.8 \pm 2.5$ &
 &
$7.8^{+a}_{-5.5}$ \\
$\chi^2 $ &&& 
0.99 & 
1.05 & 
0.40 & 
0.78 &
 &
0.93 \\
Absorbed ${F}_{X_{0.3-10.0\text{ keV}}}$ && $10^{-13}$ erg s$^{-1}$ cm$^{-2}$ &
$1.4^*$  &  
$0.29^{+0.26}_{-0.16}$ & 
$0.20^{+0.71}_{-0.17}$ &
$0.29^{+0.12}_{-0.10}$ &
 &  
$0.16^{+0.25}_{-0.14}$\\
Unabsorbed ${F}_{X_{0.3-10.0\text{ keV}}}$ && $10^{-13}$ erg s$^{-1}$ cm$^{-2}$ &
$1.4^*$ &
$0.72^{+0.51}_{-0.41}$  & 
$0.72^{+1.38}_{-0.56}$ & 
$0.76^{+0.34}_{-0.27}$  & 
 &
$0.62^{+0.96}_{-0.46} $  \\
\hline\hline
\end{tabular}
    \begin{tablenotes}
      \small
      \item $^a$These parameters are not well constrained and errors are not reported.
      \item $^*$Sample flux does not converge, values found via \texttt{calc\_energy\_flux}. \emph{XMM-Newton} fluxes are reported for pn spectra. 
      \item Spectrum on Feb 23 2017 was not able to be fit with the blurred reflection model.
    \end{tablenotes}
\end{threeparttable}}
\end{center}
\end{specialtable}  
\begin{paracol}{2}
\switchcolumn

When considering the thermal model, the temperature may be declining from January to February, with $kT$ decreasing from about 40 to 10 to 5 keV. However large errors on the best-estimate temperature at each epoch could also suggest a constant value over the time period. In the power-law model, the power-law index remains constant within error bars, with an average value around $\Gamma = 1.5$.

For completeness, the parameters for the various disk models are also given in Table~\ref{tab:ulx3_spectra}; however, error bars are unable to be calculated on many of the parameters since they are not well constrained. The column densities across models are consistently found to have decreased from 2016 to 2017, with the exception of the blurred reflection model which generally predicts higher $N_H$'s than the other models. Best-estimate values for unabsorbed model fluxes in the disk models also suggest that the source has dimmed over the duration of its observability.

The apparent dimming of ULX-3 over a period of two months raises interesting questions about its nature.  ULXs are known to be highly variable and can have flux changes over a matter of hours to days \citep{earnshawetal19, atapinetal20, pintoreetal21}. The poor $\chi^2$ in some of these fits may indicate that these models are not the best ones to use, possible if the source is not a ULX. Additional spectral modeling may help to decipher the nature of this emission.

Following \citet{Earnshaw19}, we explore various possibilities that could explain the transient nature of ULX-3.

\subsubsection{Supernova}
ULX-3 could possibly be a new supernova (SN), in which case it
exploded somewhere between 2011 and 2016. X-ray SNe can have
luminosities from 10$^{36}$ to 10$^{42}$ erg s$^{-1}$
\citep{dg12}, consistent with ULX-3's luminosity. The \emph{Chandra} spectrum of this source resembles a thermal spectrum, as has been found for the highest luminosity Type IIn SNe. We note that fits with both thermal and non-thermal
models are equally good.  The observed decline in the
flux of ULX-3 by nearly a factor of 7 over the course of two months is
faster than that seen in any known SNe
\citep{dg12}. If the source is indeed a SN, this
would suggest an extremely rapid decrease uncharacteristic of SNe. Type IIn SNe tend to decrease at a faster rate than the average SN, and SN 1986J decreases more rapidly than most SNe, at a rate $L_x \propto t^{-1.69}$ \cite{dg12}. On the other hand, the Type IIns tend to have luminosities
an order of magnitude higher than ULX-3, at least in the first 2-3
years. Even assuming that ULX-3 had a much higher luminosity earlier,
a 3 or 4 year old source decreasing by nearly a factor of 7 in two months
would make it the fastest decreasing SN ever. Thus the rapid
decrease makes it doubtful that it is a SN. Observations
at other wavelengths could help considerably in this regard.

\subsubsection{Super-Eddington Accreting Source}
The majority of ULXs may be stellar mass objects accreting at super-Eddington rates. In particular, several have been confirmed to be NSs, which can exhibit flux variability of an order of magnitude over short periods of time. A preliminary search for pulsations in the light curve of ULX-3, suggestive of NS super-Eddington accretion, proved negative. Given the various difficulties and issues in searching for pulsating sources \citep[see for instance][]{songetal20}, we have deferred a more in-depth study to a subsequent paper. A deeper \emph{Chandra} observation could assist in the search for pulsations; with the current  set of observations it is difficult to validate whether ULX-3 is a super-Eddington accretor in the ULX regime. ULX-3 lives in the lower luminosity ULX regime ($< 3 \times 10^{39} \textrm{ erg} \textrm{ s}^{-1}$). Such low luminosity sources are often ``broadened-discs'' undergoing accretion close to, and not much in excess of, the Eddington limit \citep{sutton13}.

\subsubsection{Transient Outbursts}
Another possibility for ULX-3 is that it could be an X-ray binary. A simple redshifted power-law fit with the \emph{Chandra} observation yields $\Gamma = 1.4 \pm 0.4$, typical of the sub-Eddington hard state \citep{2002A&A...390..199B, 2011MNRAS.415.2373S}. Assuming a hard-only outburst in which observed luminosities generally only reach Eddington fractions of $\sim 10\%$ \citep{2016ApJS..222...15T}, the source could be an IMBH with a mass upwards of $\sim 150 M_\odot$. However, this would imply a much cooler disk temperature than is found in the fits with the various disk models, where $T > 1 \textrm{ keV}$. 

Transient outbursts such as in X-ray binaries could explain ULX-3. The suggested time scale for the dimming of the source is consistent with the time scale of transient stellar mass black hole LMXBs
\citep{1993PASJ...45..707M, 1997ApJ...491..312C, 2012MNRAS.420.2969M}. The initial decrease in column density could be indicative of the formation and dissipation of a disk around a compact object or the onset and aftermath of Roche-lobe overflow onto the compact object in a binary system. If an optical counterpart to the X-ray outburst is observed, it is possible that ULX-3 could be interpreted as an LMXB. 

\subsubsection{Micro-tidal Disruption Events}
A micro-tidal disruption ($\mu$TDE) event occurs when a star is
disrupted and accreted onto a stellar mass black hole or IMBH. This would be a
one-time transit that would subsequently decay. The flux is found to
decay with time as $t^{-5/3}$. Typical time scales of $\mu$TDE can be
of the order of a few months, consistent with the observations of
ULX-3.  If we assume that we happened to catch the outburst just when
it happened, on day 1 (which would make the observation incredibly
well timed), then a $t^{-5/3}$ decline would mean that the source
would have declined by a factor of a 1000 over 2 months, i.e. it would
have completely disappeared. If the maximum flux was higher but
reached earlier, say even 10 days earlier, even then the decline in 2
months would be substantial. Thus, given our current data, a $\mu$TDE
could be consistent.

Unfortunately the paucity of data does not allow us to reach any
conclusions regarding ULX-3. A
second observation, with \emph{Chandra's} spatial resolution, is necessary to learn more about the nature of this intriguing
source.


\section{Conclusions}
Three ULXs have appeared over the course of nearly thirty years in the edge-on spiral galaxy NGC 891. We thus have a rare window into ULX spectral and flux evolution over an exceedingly long timescale. Previous groups have identified two of these sources (ULX-1 from \citet{1998ApJ...493..431H} in 1991 and ULX-2 from HK12 in 2011). Several groups have carried out spectral modeling of some of the individual datasets; however, none of them have modelled all the available data with a uniform set of models, and many datasets do not appear to have been studied previously. A comprehensive spectral modeling study of this region has not been carried out prior to this work. We have used the complete archival observational dataset for this region, from both \emph{XMM} and \emph{Chandra}. In the process, we have identified a third ULX that appeared in 2016 and whose flux decreased by a factor of seven over the course of two months.  Fitting the source spectra with thermal, power-law, and accretion disk models, we are unable to distinguish between the spectral fits to designate any model as being superior to the others. This is similar to HK12's findings for ULX-2. Additional multi-wavelength observations and possible timing analyses are needed in order to provide more insight into their potential nature. 

Our spectral fitting indicates that regardless of the origin of the emission, the two previously known sources are consistently ultraluminous and do not show large variability over the observational period. Long-term monitoring of ULX fluxes is rare, and this study provides evidence for consistent ultraluminous behavior over a time span of up to twenty years for these sources. Using \emph{Chandra's} high spatial resolution, we find the unabsorbed luminosity at 9 Mpc for ULX-1 to be approximately $(5.8 \pm 0.5) \times 10^{39} \textrm{ erg} \textrm{ s}^{-1}$ and for ULX-2 to be $(9 \pm 1) \times 10^{39} \textrm{ erg} \textrm{ s}^{-1}$. Assuming isotropic emission and Eddington-limited accretion, these luminosities would correspond to $\sim 50 - 80 \textrm{ M}_\odot$ black holes, placing these sources in the upper stellar black hole mass gap \citep{Woosley2021}. This would reinforce the interpretation of ULXs as simply an extension in the population of stellar mass black holes \citep{Marchant2017}. However, the high disk temperatures for these sources suggest that their spectra are not attributable to much heavier mass black holes. Multicolor disk model fitting in particular indicates that for $T \propto r^{-p}$, $p < 0.75$, suggesting that disk advection is important. Model fits for these two sources indicate exceedingly hot disk temperatures ($T_{\rm{in}} > 1 \textrm{ keV}$) which could further suggest that these two sources could be super-Eddington accretors. This would be consistent with the increasingly paradigmatic view that ULXs as a population are compact objects undergoing super-Eddington accretion \citep{Poutanen2007}. 

We also report the discovery of an additional high-energy source CXOU J022230.1 +421937 detected by \emph{Chandra} in Nov 2016. The unabsorbed luminosity of this source, which we refer to as ULX-3, is near $2 \times 10^{39} \textrm{ erg} \textrm{ s}^{-1}$, placing it on the lower boundary of the ULX luminosity range. Such a luminosity could correspond to a stellar mass black hole $\sim 20  \textrm{ M}_\odot$. Subsequent \emph{XMM} observations show that over the course of a couple of months, ULX-3 decreased by a factor of about seven in luminosity, consistent with the behavior of a variety of astrophysical transients. Follow-up spectral modeling as well as longer exposure X-ray and multi-wavelength observations could allow us to determine the nature of ULX-3 and its transitory behavior.


\vspace{6pt} 


\authorcontributions{
Conceptualization, N.E., V.V.D., and V.C.; methodology, N.E. and V.V.D.;  formal analysis, N.E., V.V.D. and V.C.; investigation, N.E., V.V.D.and V.C.; resources, V.V.D.; data curation, N.E. and V.C; writing---original draft preparation, N.E.; writing---review and editing, V.V.D. and N.E.; visualization, N.E.; supervision, V.V.D.; project administration, V.V.D.; funding acquisition, V.V.D. All authors have read and agreed to the published version of the manuscript.
}

\funding{This work is funded by NSF grant 1911061 awarded to PI Vikram V. Dwarkadas at the University of Chicago, and by the National Aeronautics and Space Administration through Chandra Award Number GO7-18066X issued by the Chandra X-ray Center, which is operated by the Smithsonian Astrophysical Observatory for and on behalf of the National Aeronautics Space Administration under contract NAS8-03060.}

\dataavailability{
Data is contained within the article. 
} 

\acknowledgments{We thank the anonymous referees for their careful reading of the paper, and informative comments and suggestions which have helped to substantially improve the paper.  We thank Franz Bauer for helpful suggestions regarding ULXs. The scientific results reported in this article are based in part on observations made by the Chandra X-ray Observatory, and in part on observations obtained with XMM-Newton, an ESA science mission with instruments and contributions directly funded by ESA Member States and NASA. The paper used data obtained from the Chandra and XMM-Newton Data Archives. This research has made use of software provided by the Chandra X-ray Center (CXC) in the application packages CIAO, ChIPS, and Sherpa, and the XMM-SAS software. }

\conflictsofinterest{The authors declare no conflict of interest. The funders had no role in the design of the study; in the collection, analyses, or interpretation of data; in the writing of the manuscript, or in the decision to publish the~results.} 

\end{paracol}
\reftitle{References}


\externalbibliography{yes}
\bibliography{ulxpapers}

\end{document}